\documentclass[longauth]{aa}
\usepackage[utf8]{inputenc}

\usepackage{savesym}
\usepackage{amsmath}
\savesymbol{iint}
\usepackage[varg]{txfonts}
\restoresymbol{TXF}{iint}
\usepackage{graphicx}
\usepackage{units}
\usepackage[breaklinks, colorlinks, citecolor=blue, linkcolor=blue]{hyperref}
\usepackage{threeparttable} 
\usepackage{siunitx}
\usepackage{natbib}
\usepackage{blindtext}
\usepackage{longtable}
\usepackage{pdflscape} 
\usepackage{float}
\usepackage[caption = false]{subfig}
\usepackage[multiple]{footmisc}

\bibpunct{(}{)}{;}{a}{}{,} 

\let\orgautoref\autoref
\renewcommand{\autoref}
        {\def\equationautorefname{Eq.}%
         \def\figureautorefname{Fig.}%
         \def\sectionautorefname{Sect.}%
         \def\subsectionautorefname{Sect.}%
         \def\subsubsectionautorefname{Sect.}%
         \orgautoref}

\newcommand*\samethanks[1][\value{footnote}]{\footnotemark[#1]}

\newcommand{\tx}[1]{\mathrm{#1}}

\makeatletter

\def\instrefs#1{{\def\scsep{\def\scsep{,}}\@for\w:=#1\do{\scsep\ref{inst:\w}}}}
\renewcommand{\inst}[1]{\unskip$^{\instrefs{#1}}$}

\DeclareSIUnit\parsec{pc}
\DeclareSIUnit\as{arcsec}
\DeclareSIUnit\pixel{pix}
\DeclareSIUnit\year{yr}
\usepackage{color}

\begin{document}

\title{Discovery of a hot, transiting, Earth-sized planet and a second temperate, non-transiting planet around the M{4} dwarf {GJ~3473} (TOI-488)\thanks{RV data are only available in electronic form
at the CDS via anonymous ftp to cdsarc.u-strasbg.fr (130.79.128.5)
or via \url{http://cdsweb.u-strasbg.fr/cgi-bin/qcat?J/A+A/}}}
\titlerunning{Discovery of a planetary system around {GJ~3473}}

\author{J.~Kemmer\inst{lsw}\thanks{Fellow of the International Max Planck Research School for Astronomy and Cosmic Physics at the University of Heidelberg (IMPRS-HD).}
\and S.~Stock\inst{lsw}\samethanks
\and D.~Kossakowski\inst{mpia}\samethanks
\and A.~Kaminski\inst{lsw}
\and K.~Molaverdikhani\inst{lsw,mpia}
\and M.~Schlecker\inst{mpia}\samethanks
\and J.\,A.~Caballero\inst{cabesac}
\and P.\,J.~Amado\inst{iaa} 
\and N.~Astudillo-Defru\inst{ucsc} 
\and X.~Bonfils\inst{grenoble} 
\and D.~Ciardi\inst{caltech} 
\and K.\,A.~Collins\inst{cfa} 
\and N.~Espinoza\inst{stsci} 
\and A.~Fukui\inst{utdep,ull} 
\and T.~Hirano\inst{tit} 
\and J.\,M.~Jenkins\inst{ames} 
\and D.\,W.~Latham\inst{cfa} 
\and E.\,C.~Matthews\inst{kavli} 
\and N.~Narita\inst{komaba,jst,aco,iac} 
\and E.~Pall\'e\inst{iac,ull}
\and H.~Parviainen\inst{iac,ull}
\and A.~Quirrenbach\inst{lsw}
\and A.~Reiners\inst{iag}
\and I.~Ribas\inst{ice,ieec}
\and G.~Ricker\inst{kavli}
\and J.\,E.~Schlieder\inst{nasa}
\and S.~Seager\inst{kavli,mit1,mit2}
\and R.~Vanderspek\inst{kavli}
\and J.\,N.~Winn\inst{prince} 
\and J.\,M.~Almenara\inst{grenoble}
\and V.\,J.\,S.~B\'ejar\inst{iac,ull}
\and P.~Bluhm\inst{lsw}\samethanks
\and {F.~Bouchy\inst{ogu}}
\and P.~Boyd\inst{nasa}
\and J.\,L.~Christiansen\inst{caltech}
\and C.~Cifuentes\inst{cabesac}
\and R.~Cloutier\inst{cfa}
\and K.\,I.~Collins\inst{gmu}
\and M.~Cort\'es-Contreras\inst{cabesac}
\and I.\,J\,M.~Crossfield\inst{kavli}
\and N.~Crouzet\inst{esa}
\and J.\,P.~de Leon\inst{utda}
\and {D.\,D.~Della-Rose}\inst{usafa}
\and X.~Delfosse\inst{grenoble}
\and S.~Dreizler\inst{iag}
\and E.~Esparza-Borges\inst{ull} 
\and Z.~Essack\inst{mit1,mit2}
\and Th.~Forveille\inst{grenoble}
\and P.~Figueira\inst{eso,caup}
\and D.~Galad\'{\i}-Enr\'{\i}quez\inst{caha}
\and T.~Gan\inst{tu}
\and A.~Glidden\inst{mit1,kavli}
\and E.~J.~Gonzales\inst{dauc,nsfg}
\and P.~Guerra\inst{oaa}
\and H.~Harakawa\inst{subaru}
\and A.\,P.~Hatzes\inst{tls}
\and Th.~Henning\inst{mpia}
\and E.~Herrero\inst{ieec}
\and K.~Hodapp\inst{hawaii}
\and Y.~Hori\inst{aco,nao}
\and S.\,B.~Howell\inst{ames}
\and M.~Ikoma\inst{utdep}
\and K.~Isogai\inst{kyoto}
\and S.\,V.~Jeffers\inst{iag}
\and M.~K\"urster\inst{mpia}
\and K.~Kawauchi\inst{utdep}
\and T.~Kimura\inst{utdep}
\and P.~Klagyivik\inst{iac,ull,dlr}
\and T.~Kotani\inst{aco,nao,guas}
\and T.~Kurokawa\inst{nao,tuat}
\and N.~Kusakabe\inst{aco,nao}
\and M.~Kuzuhara\inst{aco,nao}
\and M.~Lafarga\inst{ice,ieec}
\and J.\,H.~Livingston\inst{utda}
\and R.~Luque\inst{iac,ull}
\and R.~Matson\inst{usno}
\and J.\,C.~Morales\inst{ice,ieec}
\and M.~Mori\inst{utda}
\and P.\,S.~Muirhead\inst{bou}
\and F.~Murgas\inst{iac,ull}
\and J.~Nishikawa\inst{nao,aco,guas}
\and T.~Nishiumi\inst{guas,nao}
\and M.~Omiya\inst{aco,nao}
\and S.~Reffert\inst{lsw}
\and C.~Rodr\'iguez L\'opez\inst{iaa}
\and N.\,C.~Santos\inst{caup,dfap}
\and P.~Sch\"ofer\inst{iag}
\and R.\,P.~Schwarz\inst{pata}
\and B.~Shiao\inst{stsci}
\and M.~Tamura\inst{utda,aco,nao}
\and Y.~Terada\inst{utda}
\and J.\,D.~Twicken\inst{ames,seti}
\and A.~Ueda\inst{nao,aco,guas}
\and S.~Vievard\inst{subaru}
\and N.~Watanabe\inst{guas,aco,nao}
\and M.~Zechmeister\inst{iag}
}

\institute{
\label{inst:lsw}Landessternwarte, Zentrum f\"ur Astronomie der Universit\"at Heidelberg, K\"onigstuhl 12, 69117 Heidelberg, Germany \email{jkemmer@lsw.uni-heidelberg.de}
\and
\label{inst:mpia}Max-Planck-Institut f\"ur Astronomie, K\"onigstuhl 17, 69117 Heidelberg, Germany
\and
\label{inst:cabesac}Centro de Astrobiolog\'ia (CSIC-INTA), ESAC, Camino bajo del castillo s/n, 28692 Villanueva de la Ca\~nada, Madrid, Spain
\and
\label{inst:iaa}Instituto de Astrof\'isica de Andaluc\'ia (IAA-CSIC), Glorieta de la Astronom\'ia s/n, 18008 Granada, Spain
\and
\label{inst:ucsc}Departamento de Matem\'atica y F\'is\'ica Aplicadas, Universidad Cat\'olica de la Sant\'isima Concepci\'on, Alonso de Rivera 2850, Concepci\'on, Chile
\and
\label{inst:grenoble}Univ. Grenoble Alpes, CNRS, IPAG, F-38000 Grenoble, France
\and
\label{inst:caltech}Caltech/IPAC-NASA Exoplanet Science Institute, 770 S. Wilson Avenue, Pasadena, CA 91106, USA
\and
\label{inst:cfa}Center for Astrophysics \textbar \ Harvard \& Smithsonian, 60 Garden Street, Cambridge, MA 02138, USA
\and
\label{inst:stsci}Space Telescope Science Institute, 3700 San Martin Drive, Baltimore, MD 21218, USA
\and
\label{inst:utdep}Department of Earth and Planetary Science, Graduate School of Science, The University of Tokyo, 7-3-1 Hongo, Bunkyo-ku, Tokyo 113-0033, Japan
\and 
\label{inst:ull}Departamento de Astrof\'isica, Universidad de La Laguna, 38206 La Laguna, Tenerife, Spain
\and
\label{inst:tit}Department of Earth and Planetary Sciences, Tokyo Institute of Technology, 2-12-1 Ookayama, Meguro-ku, Tokyo 152-8551, Japan
\and
\label{inst:ames}NASA Ames Research Center, Moffett Field, CA 94035, USA
\and
\label{inst:kavli}Department of Physics and Kavli Institute for Astrophysics and Space Research, Massachusetts Institute of Technology, Cambridge, MA 02139, USA
\and
\label{inst:komaba}Komaba Institute for Science, The University of Tokyo, 3-8-1 Komaba, Meguro, Tokyo 153-8902, Japan
\and
\label{inst:jst}JST, PRESTO, 3-8-1 Komaba, Meguro, Tokyo 153-8902, Japan
\and
\label{inst:aco}Astrobiology Center, 2-21-1 Osawa, Mitaka, Tokyo 181-8588, Japan
\and 
\label{inst:iac}Instituto de Astrof\'isica de Canarias (IAC), 38205 La Laguna, Tenerife, Spain
\and
\label{inst:iag}Institut f\"ur Astrophysik, Georg-August-Universit\"at, Friedrich-Hund-Platz 1, 37077 G\"ottingen, Germany
\and 
\label{inst:ice}Institut de Ci\`encies de l’Espai (ICE, CSIC), Campus UAB, Can Magrans s/n, 08193 Bellaterra, Spain
\and
\label{inst:ieec}Institut d’Estudis Espacials de Catalunya (IEEC), 08034 Barcelona, Spain
\and
\label{inst:nasa}NASA Goddard Space Flight Center, 8800 Greenbelt Road, Greenbelt, MD 20771, USA
\and
\label{inst:mit1}Department of Earth, Atmospheric and Planetary Sciences, Massachusetts Institute of Technology, Cambridge, MA 02139, USA
\and
\label{inst:mit2}Department of Aeronautics and Astronautics, MIT, 77 Massachusetts Avenue, Cambridge, MA 02139, USA
\and
\label{inst:prince}Department of Astrophysical Sciences, Princeton University, 4 Ivy Lane, Princeton, NJ 08544, USA
\and
\label{inst:ogu}Observatoire de Gen\`eve, Universit\'e de Gen\`eve, 51 ch. des Maillettes, 1290 Sauverny, Switzerland
\and
\label{inst:gmu}George Mason University, 4400 University Drive, Fairfax, VA, 22030 USA
\and
\label{inst:esa}European Space Agency, ESTEC, Keplerlaan 1, 2201 AZ Noordwijk, The Netherlands
\and
\label{inst:utda}Department of Astronomy, The University of Tokyo, 7-3-1 Hongo, Bunkyo-ku, Tokyo 113-0033, Japan
\and
\label{inst:usafa}{Department of Physics, United States Air Force Academy, Colorado, CO 80840, USA}
\and
\label{inst:eso}European Southern Observatory, Alonso de C\'ordova 3107, Vitacura, Regi\'on Metropolitana, Chile
\and
\label{inst:caup}Instituto de Astrof\'isica e Ci\^encias do Espa\c{c}o, Universidade do Porto, CAUP, Rua das Estrelas, PT4150-762 Porto, Portugal
\and
\label{inst:caha}Centro Astron\'omico Hispano-Alem\'an, Observatorio de Calar Alto, Sierra de los Filabres, 04550 G\'ergal, Spain
\and
\label{inst:tu}Department of Astronomy, Tsinghua University, Beijing 100084, China
\and
\label{inst:dauc}Department of Astronomy and Astrophysics, University of California, Santa Cruz, 1156 High St. Santa Cruz , CA 95064, USA
\and
\label{inst:nsfg}National Science Foundation Graduate Research Fellow
\and
\label{inst:oaa}Observatori Astron\'omic Albany\'a, Camí de Bassegoda S/N, Albany\'a 17733, Girona, Spain
\and
\label{inst:subaru}Subaru Telescope, 650 N. Aohoku Place, Hilo, HI 96720, USA
 \newpage
\and 
\label{inst:tls}Th\"uringer Landessternwarte Tautenburg, Sternwarte 5, 07778 Tautenburg, Germany
\and
\label{inst:hawaii}University of Hawaii, Institute for Astronomy, 640 N. Aohoku Place, Hilo, HI 96720, USA
\and
\label{inst:nao}National Astronomical Observatory of Japan, 2-21-1 Osawa, Mitaka, Tokyo 181-8588, Japan
\and
\label{inst:kyoto}Okayama Observatory, Kyoto University, 3037-5 Honjo, Kamogatacho, Asakuchi, Okayama 719-0232, Japan
\and
\label{inst:dlr}Deutsches Zentrum f\"ur Luft und Raumfahrt, Institut f\"ur Planetenforschung Rutherfordstrasse 2, 12489 Berlin, Germany, DE
\and
\label{inst:guas}Department of Astronomical Science, The Graduated University for Advanced Studies, SOKENDAI, 2-21-1, Osawa, Mitaka, Tokyo, 181-8588 Japan
\and
\label{inst:tuat}Tokyo University of Agriculture and Technology, 3-8-1, Saiwai-cho, Fuchu, Tokyo, 183-0054, Japan
\and
\label{inst:usno}U.S. Naval Observatory, 3450 Massachusetts Avenue NW, Washington, D.C. 20392, USA
\and
\label{inst:bou}Department of Astronomy, Institute for Astrophysical Research, Boston University, 725 Commonwealth Ave. Room 514, Boston, MA 02215, USA
\and
\label{inst:dfap}Departamento de F\'isica e Astronomia, Faculdade de Ci\^encias, Universidade do Porto, Rua do Campo Alegre, 4169-007 Porto, Portugal
\and
\label{inst:pata}Patashnick Voorheesville Observatory, Voorheesville, NY 12186, USA
\and
\label{inst:seti}SETI Institute, Mountain View, CA 94043, USA
}

\date{Received dd July 2020 /
Accepted <day month year>}

\abstract
{We present the confirmation and characterisation of {GJ~3473}~b ({G~50--16}, TOI-488.01), a hot Earth-sized planet {orbiting an M4 dwarf star}, whose transiting signal ($P=\SI{1.1980035\pm0.0000018}{d}$) was first detected by the {\em Transiting Exoplanet Survey Satellite} ({\em TESS}). 
{Through} a joint modelling of follow-up radial velocity observations with CARMENES, IRD, and HARPS together with extensive ground-based photometric follow-up observations with LCOGT, MuSCAT, and MuSCAT2, we determined a precise planetary mass, $M_b = \SI{1.86\pm0.30}{M_\oplus,}$ and radius, $R_b = \SI{1.264\pm0.050}{R_\oplus}$. Additionally, we report the discovery of a second, temperate, non-transiting planet in the system, {GJ~3473}~c, which has a minimum mass, $M_c \sin{i} = \SI{7.41\pm0.91}{M_\oplus,}$ and orbital period,  $P_c=\SI{15.509\pm0.033}{d}$. The inner planet of the system, {GJ~3473}~b, is one of the hottest transiting Earth-sized planets known thus far, accompanied  by a {dynamical} mass measurement, which makes it a particularly {attractive} target for {thermal} emission spectroscopy.}

\maketitle

\section{Introduction}

The detection of transiting planets with the radial velocity (RV) method enables us to derive a comprehensive characterisation of their properties. In particular, {it permits the measurement of} a {dynamical} planetary mass and, hence, a measurement of the planetary mean density {when combined with the planetary radius derived from the transit light curve}. From comparisons with theoretical models, the density of a planet provides information about its composition and structure and, therefore, it plays a key role in understanding planet formation {and evolution} \citep[e.g.][]{Southworth.2010, Marcy.2014, Rogers.2015, Fulton.2017, Bitsch.2019, Zeng.2019}.
Furthermore, additional non-transiting planets in the system can be detected with the RV method. Such multi-planetary systems hold valuable information because the dynamical interaction between the planets can have a significant influence {on} their formation and evolution, as well as shaping the currently observed architecture of the system \citep[e.g.][]{Lissauer.2007, Zhu.2012, AngladaEscude.2013, Mills.2017, Morales.2019}.

A significant fraction of the over \num{3000} transiting exoplanets known today\footnote{On 26 August 2020, 3189 transiting exoplanets were listed by \url{exoplanetarchive.ipac.caltech.edu/}} were discovered by the \emph{Kepler} satellite \citep{Borucki.2010, Borucki.2016}. However, \emph{Kepler's} focus on faint stars ($K_p {>} \SI{12}{mag}$) impeded detailed follow-up studies of those planets using ground-based facilities. In contrast, the {\em Transiting Exoplanet Survey Satellite} \citep[\emph{TESS;}][]{Ricker.2015} is now filling in this gap. To date, \emph{TESS} has already found more than 50 confirmed transiting planets, and many more candidates, orbiting bright, nearby stars ($G \sim$ 6--13\,mag, $d \sim$ 10--340\,pc). One of its level-one science requirements is to measure the masses for 50 transiting planets with radii smaller than \SI{4}{R_\oplus} by RV follow-up observations\footnote{\url{https://heasarc.gsfc.nasa.gov/docs/tess/primary-science.html}, visited on 28 June 2020}.
What is particularly interesting in this regime are planets that are orbiting M dwarf stars. The relative transit depth, and thus the detection probability of rocky planets around M dwarfs, is much higher compared to larger stars of earlier spectral types. Still, despite M dwarfs being the most common stars in our Galaxy \citep[e.g.][]{Chabrier.2003, Henry.2006} and the fact that small planets are more abundant around later type stars \citep{Howard.2012, Bonfils.2013, Mulders.2015, Dressing.2015, Gaidos.2016, HardegreeUllman.2019}, only a few precise {dynamical} masses of such planets have currently been determined. 
Prior to the \emph{TESS} mission, only \num{12} planets with radii smaller than $R = \SI{2}{R_\oplus}$ and {dynamical} mass measurements to a precision better than \SI{30}{\percent} were known to orbit stars with temperatures, $T_{\rm eff}<\SI{4000}{\kelvin}$. Thanks to the intensive RV follow-up of \emph{TESS} planet candidates, this number {already} increased by seven new planets (see \autoref{tab:comparative_planets} for the full list).
{The brightness of these cool \emph{TESS} host stars, combined with their small size, makes many of them ideal targets for atmospheric characterisation by transmission or thermal emission spectroscopy with upcoming space-borne or ground-based instruments \citep{Kempton.2018, Batalha.2018}.}

Here, we report the discovery of a planetary system around the intermediate M dwarf \object{{GJ~3473}}. The inner, Earth-sized planet was first detected as a transiting planet candidate by \emph{TESS}. Our extensive RV monitoring campaign, using CARMENES, IRD, and HARPS, confirms its planetary nature and reveals a second, more massive, non-transiting planet {on a wider orbit}. This paper  is structured as follows: \autoref{sec:data} describes the data used in this study. In \autoref{sec:stellar_prop}, the properties of the host star are presented. The analysis of the data is set out in \autoref{sect:analysis} and the results are discussed in \autoref{sec:discussion}. Finally, we give our conclusions in \autoref{sec:conclusions}.%

\section{Data}%
\label{sec:data}%

\subsection{\emph{TESS}}%

\begin{figure}
\centering
\includegraphics[]{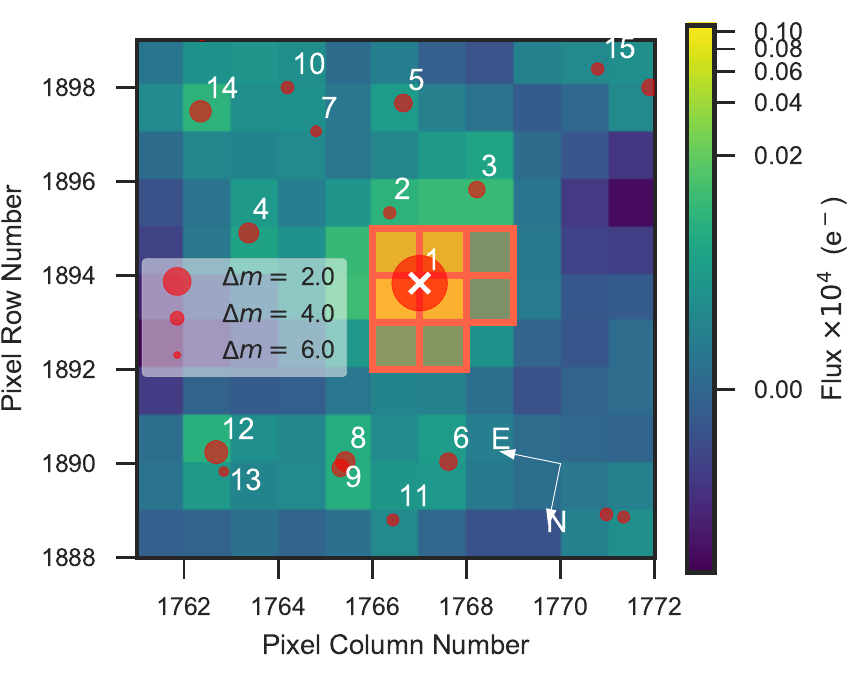}
\caption{\emph{TESS} TPF of {GJ~3473}. 
The planet-host star is marked by a white cross and the  {pixels of the aperture mask used for the retrieval of the light curve are highlighted with orange borders.} Sources listed in the Gaia DR2 catalogue \citep{GaiaCollaboration.2018} are indicated by red circles (size proportional to their brightness {difference with {GJ~3473}}).
Source \#3 is LP~544--12, the common proper motion companion to {GJ~3473}.}
\label{fig:tpf_plot}
\end{figure}

\begin{figure*}
    \centering
    \includegraphics[]{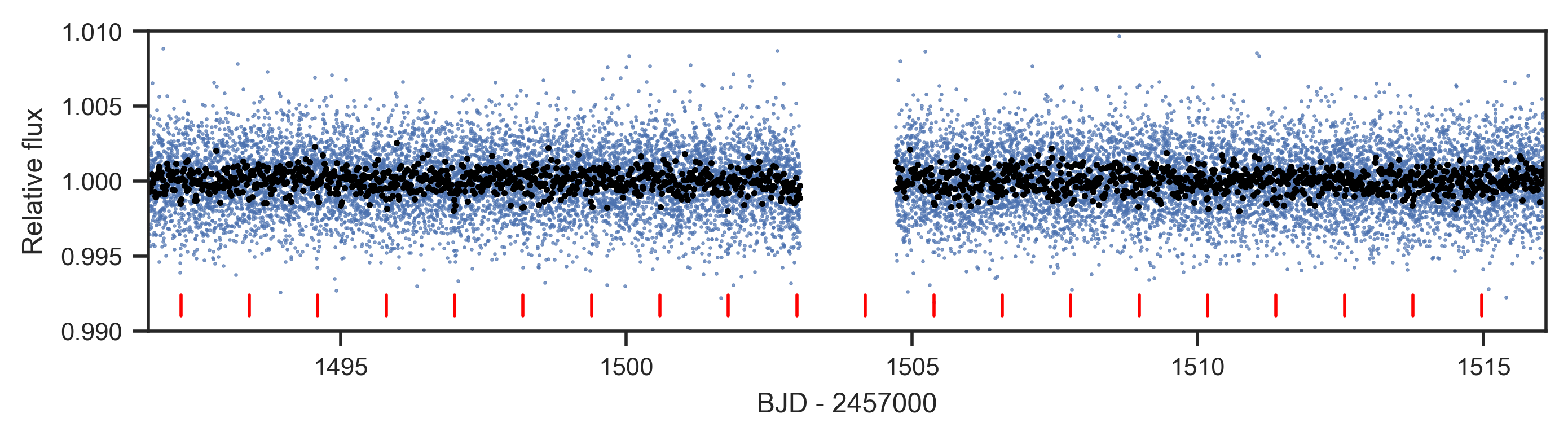}
    \caption{\emph{TESS} systematic-corrected PDC-SAP light curve. {The blue dots are the measurements and the black dots are the data binned to \SI{20}{\minute}.}  The transit times are marked by red ticks.}
    \label{fig:tess_lc}
\end{figure*}

{GJ~3473} {(TIC\,452866790)} was observed by \emph{TESS} with a two-minute cadence in Sector 7 (Camera \#1, CCD \#3) between 7 January and 2 February 2019 {and is listed to have a transiting planet candidate on the \emph{TESS} releases website (TOI--488.01)}. Due to its proximity to the ecliptic plane, it will not be observed again by \emph{TESS} during its primary mission, but will be revisited in Sector 34 of the {\emph{TESS} extended mission in the third year}\footnote{\url{https://heasarc.gsfc.nasa.gov/cgi-bin/tess/webtess/wtv.py?Entry=452866790}, visited on 28 April 2020}. 
The time series had a gap between $\mathrm{BJD} = 2458503.04$ and $\mathrm{BJD} = 2458504.71$ because of the data downlink and telescope re-pointing {(see \autoref{fig:tess_lc})}. 
The light curves produced by the Science Processing Operations Center \citep[SPOC;][]{Jenkins.2016} are available on the Mikulski Archive for Space Telescopes\footnote{\url{https://mast.stsci.edu}}. 
For our analysis, we used the systematics-corrected simple aperture photometry (PDC-SAP) light curve \citep{Smith.2012, Stumpe.2012, Stumpe.2014}.
A plot of the target pixel file (TPF) and the aperture mask that is used for the simple aperture photometry (SAP), generated with {\tt tpfplotter}\footnote{\url{https://github.com/jlillo/tpfplotter}},
is shown in \autoref{fig:tpf_plot}.
The {\em TESS} data have a median internal uncertainty of \SI{2.35}{ppt} (parts per thousand) and root mean square (rms) of \SI{2.2}{ppt} around the mean. See \cite{Luque.2019}, \cite{Dreizler.2020}, \cite{Nowak.2020}, and \cite{Bluhm.2020} for further details {on the applied methodology}.

\subsection{High-resolution spectroscopy}
High-resolution follow-up spectroscopy of the \emph{TESS} planet candidates is arranged by the \emph{TESS} follow-up programme (TFOP), `Precise Radial Velocities' SG4 subgroup\footnote{\url{https://tess.mit.edu/followup/}}. The goal is to achieve a full validation of the candidates and {to} ultimately provide {their} mass measurement.
\paragraph{CARMENES.}
As part of the CARMENES guaranteed time observation programme to search for exoplanets around M dwarfs \citep{Reiners.2018}, we observed {GJ~3473} with CARMENES \citep[Calar Alto high-Resolution search for M dwarfs with Exoearths with Near-infrared and visible Echelle Spectrographs;][]{Quirrenbach.2014}.
CARMENES is a high-resolution spectrograph at the \SI{3.5}{m} Calar Alto telescope that consists of two cross-dispersed echelle channels operating in the spectral ranges of \SIrange{0.52}{0.96}{\micro \meter} in the visible light (VIS, $R = \num{94600}$)  and \SIrange{0.96}{1.71}{\micro \meter} in the near infrared (NIR, $R = \num{80400}$), respectively. The observations began at the end of March 2019, just after the announcement of the transiting planet candidate, and ended in January 2020. In this period, we collected 67 pairs of VIS and NIR spectra with exposure times of about \SI{30}{min} each. Within the standard CARMENES data flow, the spectra are calibrated using \texttt{CARACAL} \citep{Caballero.2016b}, while the {RVs} are calculated using \texttt{SERVAL} \citep{Zechmeister.2018}. {The RVs are} corrected for barycentric motion, secular perspective acceleration, as well as instrumental drift. To reconstruct small systematic radial-velocity variations, so called nightly zero-point offsets, we use the measured RVs of all other stars with only small intrinsic RV variations from the respective observing nights \citep[see][for details]{Trifonov.2018, TalOr.2019, Trifonov.2020}. Spectra without simultaneous Fabry-P\'erot drift measurements or a signal-to-noise ratio (S/N) lower than \num{10} are excluded during the process, which results in a total of 64 RV measurements in the VIS and 66 in the NIR. The RVs show a median internal uncertainty of \SI{2.1}{\meter \per \second} and {a} weighted rms (wrms) of \SI{3.8}{\meter \per \second} in the VIS and \SI{11.7}{\meter \per \second} and \SI{15.6}{\meter \per \second} in the NIR, respectively. {The high scatter in the NIR channel corresponds to our expectation from the photon-noise limit considering the median measured S/N of $\sim 63$ for the NIR observations \citep[see][for a detailed analysis of the performance of CARMENES]{Bauer.2020}. Due to the low RV amplitude of the transiting planet candidate ($K \approx \SI{2.2}{\meter\per\second}$), we therefore used only the VIS data for this study.}

\paragraph{IRD.}
In the course of the Subaru IRD {{\emph{TESS}} Intensive Follow-up Project (proposal S19A-069I)}, 
we observed {GJ~3473} with the InfraRed Doppler spectrograph \citep[IRD;][]{Kotani.2018}, 
a near-infrared, adaptive-optics (AO) corrected, high-resolution spectrograph (\SIrange{0.97}{1.75}{\micro \meter}, $R\approx \num{70000}$) installed on
the Subaru 8.2\,m telescope.
The integration time was set to \SIrange{300}{600}{\second} so that the extracted
one-dimensional spectra have S/N ratios of \numrange[range-phrase=--]{50}{70} per pixel at \SI{1000}{nm}. 
A total of 56 frames were acquired for {GJ~3473}  by IRD on 12 different nights between April 2019 and December 2019, all of which had simultaneous reference spectra of the laser frequency comb. The reduction of the raw data was performed with the \texttt{IRAF} echelle package \citep{Tody.1993}, including the wavelength calibration using thorium-argon hollow cathode lamps. 
For the RV analyses, wavelengths were re-calibrated more precisely based on the laser frequency comb spectra. RVs were measured using the forward 
modelling technique described by \citet{Hirano.2020}, in which the time-variable telluric absorptions and the instantaneous instrumental profile of the spectrograph were modelled and taken into account in the RV fits. The IRD RVs show a median internal uncertainty of \SI{4.1}{\meter\per\second} and {a wrms} of \SI{8.0}{\meter\per\second}.

\paragraph{HARPS.}
{GJ~3473} was also observed by the
High Accuracy Radial velocity Planet Searcher \citep[HARPS;][]{Mayor.2003} {as part of the ESO programme 1102.C-0339(A)}. The spectrograph, installed at the ESO La Silla 3.6\,m telescope, covers the spectral range from \SIrange{0.378}{0.691}{\micro \meter} and has a resolution of $R= \num{110000}$.  The \num{32} observations presented here were taken between May 2019 and March 2020. Their exposure times ranged between \SIrange[range-phrase=\text{ and } ]{30}{40}{min}. We use the reduced spectra from the HARPS Data Reduction Software \citep[{\texttt{DRS};}][]{Lovis.2007} and compute their RVs following \citet{AstudilloDefru.2017c}{, which resulted in a lower rms scatter compared to the RVs retrieved with the \texttt{SERVAL} pipeline.} They are calibrated for the barycentric motion, secular perspective acceleration, and instrumental drift. For the HARPS RVs, we obtain a median internal uncertainty of \SI{3.4}{\meter\per\second} and a wrms of \SI{4.8}{\meter\per\second}.

\subsection{Ground-based transit follow-up}
\label{subsec:data_transit}
The TFOP subgroup SG1 provides seeing-limited photometry follow-up observations of the \emph{TESS} planet candidates in order to supplement the available photometry and to provide improved {ephemerides} for the targets. An overview of the observations, the instruments and the filters used is given in \autoref{tab:phot_follwup}.

\begin{table*}
    \centering
    \caption{Summary of the ground-based transit follow-up observations.} 
    \label{tab:phot_follwup}
    \begin{tabular}{l
                    c
                    S[table-format=3]
                    c
                    S[table-format=3]
                    S[table-format=3]
                    S[table-format=3]
                    S[table-format=2]
                    S[table-format=1.2]}
        \hline\hline
        \noalign{\smallskip}
        Telescope  & Date       & {Transit}\tablefootmark{(a)}  & Filter & {t$_\mathrm{exp}$} & {Duration\tablefootmark{(b)}} & {N$_\mathrm{obs}$} & {Aperture}  & {rms} \\
                   &            &            &        & {[\si{\second}]}                  & {[min]}            &                  & {[pix]}                       & {[ppt]}   \\
        \noalign{\smallskip}
        \hline
        \noalign{\smallskip}
       LCOGT McD  & 2019-03-19 & 58         & $z_s$  & 100                    & 234                & 110              & 18                            & 1.25       \\
        MuSCAT2    & 2019-12-21 & 290        & $i$    & 30                     & 237                & 675              & 32                            & 1.75       \\
        MuSCAT2    & 2019-12-21 & 290        & $z_s$  & 20                     & 237                & 457              & 32                            & 1.65       \\
        MuSCAT2    & 2020-01-02 & 300        & $r$    & 18                     & 254                & 823              & 32                            & 2.45       \\
        MuSCAT2    & 2020-01-02 & 300        & $i$    & 18                     & 254                & 845              & 32                            & 2.12       \\
        MuSCAT2    & 2020-01-02 & 300        & $z_s$  & 18                     & 254                & 845              & 32                            & 1.60       \\
        MuSCAT     & 2020-01-18 & 313        & $r$    & 20                     & 202                & 553              & 24                            & 1.85       \\
        MuSCAT     & 2020-01-18 & 313        & $z_s$  & 20                     & 202                & 551              & 26                            & 1.11       \\
        LCOGT CTIO & 2020-02-21 & 341        & $i_p$  & 60                     & 224                & 145              & 20                            & 1.56       \\
        LCOGT CTIO & 2020-02-27 & 346        & $i_p$  & 60                     & 229                & 145              & 19                            & 1.58       \\
        LCOGT SAAO & 2020-03-13 & 359        & $z_s$  & 100                    & 230                & 101              & 16                            & 1.10       \\
        \noalign{\smallskip}
        \hline
    \end{tabular}
    \tablefoot{
        \tablefoottext{a}{Transit number after the first transit observed by \emph{TESS}.}
        \tablefoottext{b}{Time-span of the observation.}}
\end{table*}

\paragraph{LCOGT.}
We used four transit observations of {GJ~3473} from the Las Cumbres Observatory global telescope network \citep[LCOGT;][]{Brown.2013}. The observations were 
taken with the SINISTRO CCDs at the \SI{1}{\meter} telescopes of the
LCOGT, which have a pixel scale of \SI{0.389}{arcsec/pix} and a field of view (FOV) of $\SI{26}{arcmin}\times \SI{26}{arcmin}$ each. The first transit was observed from the McDonald Observatory (McD) on 19 March 2019 in the $z_s$ filter, two transits were observed from the Cerro Tololo Interamerican Observatory (CTIO) on 21 and 27 February 2020 in $i_p$ filter and one transit in $z_s$ filter on 13 March 2020 from South African Astronomical Observatory (SAAO). We calibrated the images with the standard LCOGT \texttt{Banzai} pipeline \citep{McCully.2018} and extracted the light curves using \texttt{AstroImageJ} \citep{Collins.2017}. 

\paragraph{MuSCAT.}
{GJ~3473} was observed on 18 January 2020 by the Multi-color Simultaneous Camera for studying Atmospheres of Transiting planets  \citep[MuSCAT;][]{Narita.2015} mounted at the \SI{1.88}{\meter} telescope at the {Okayama Astro-Complex on Mt. Chikurinji, Japan}. MuSCAT is a multi-colour instrument that performs imaging in the $g$, $r$ and $z_s$-filter bands at the same time. Each camera has a FOV of $\SI{6.1}{arcmin}\times\SI{6.1}{arcmin}$ with a pixel scale of \SI{0.358}{arcsec/pix}. Due to a large scatter in the $g$ band, we only use the $r$ and $z_s$ light curves here. The individual images are corrected for dark current and flat fields{,} and the light curves are generated using a custom pipeline that is described in \cite{Fukui.2011}. 

\paragraph{MuSCAT2.}
We made use of two transit observations from MuSCAT2 \citep{Narita.2019}. The instrument is mounted at the \SI{1.52}{m} Telescopio Carlos S\'anchez at the Observatorio del Teide, Spain. MuSCAT2 operates simultaneously in the $g$, $r$, $i$, and $z_s$ passbands and has a FOV of $\SI{7.4}{arcmin}\times\SI{7.4}{arcmin}$ at \SI{0.44}{arcsec/pix} resolution. One transit was observed on 21 December 2019, from which we use the observations in the $i$ and $z_s$ bands. The other transit was observed on 2 January 2020, from which we use the observations in the $r$, $i,$ and $z_s$ bands. Both transits were observed defocussed to {optimise} the quality of the photometry. The transit signal had too low S/N in the $g$ band to be useful in the fitting, and the $r$ band observations were affected by systematics on the night of 21 December. The photometry was produced using a dedicated MuSCAT2 photometry pipeline \citep[see][for details]{Parviainen.2019} and the detrended light curves were created by a fit that aims to simultaneously choose the best target and comparison star apertures, model the systematics using a linear term, and include the transit using \texttt{PyTransit} \citep{Parviainen.2015}.

\paragraph{USAFA.}
{We used the brand-new, recently commissioned 1\,m USAFA Telescope, which is an optically-fast $f/6$ Ritchey-Chr\'etien telescope with a wide field of view 0.9\,deg$^2$ and an STA1600 CCD installed on the outskirts of Colorado Springs.
We observed {GJ~3473} on 04 March 2020.
The USAFA data did not firmly detect the transit on target, but ruled out nearby eclipsing binaries in all other stars within the apertures of {\em TESS}, LCOGT, and MuSCAT/2 (Fig.~\ref{fig:tpf_plot}).}

\subsection{Photometric monitoring}
We {used long-term} photometric monitoring of {GJ~3473} to search for periodic signals associated with the rotation period of the star. 

\paragraph{TJO.}
We observed {GJ~3473} with the 80\,cm Joan Or\'o telescope (TJO) at Observatori Astron\`omic del Montsec, Spain. The star was monitored between 31 January and 8 May 2020 for a total of 32 nights. 
Our observations were {performed} in the Johnson $R$ filter by using the main imaging camera LAIA, which has a 4k\,$\times$\,4k back illuminated CCD with a pixel scale of 0.4\,arcsec and a FOV of 30\,arcmin.  We calibrated each image for bias and dark current as well as applied flat field images using the \texttt{ICAT} pipeline \citep{Colome.2006}. Differential photometry was extracted with \texttt{AstroImageJ} using the aperture size and set of comparison stars that {minimised} the rms of the photometry. Low S/N data due to high airmass or bad weather were removed. The data were binned to one measurement per hour. The median internal uncertainty is \SI{2.7}{ppt}{,} while the rms is \SI{9.4}{ppt} around the mean. For the estimation of the stellar rotation period with a Gaussian process{,} we binned these data to one data point per night. This reduces short term variations caused by jitter and yields a median internal uncertainty of \SI{2.9}{ppt} and a rms of \SI{7.4}{ppt} around the mean.

\paragraph{MEarth.}
The all-sky transit survey MEarth consists of 16 robotic \SI{40}{cm} telescopes with a FOV of \SI{26}{arcmin\squared} {located at} two observatories in the southern and northern hemisphere \citep{Berta.2012}. We use archival photometric monitoring data from the Mearth-North project DR8\footnote{\tt \url{https://www.cfa.harvard.edu/MEarth/DR8/}} covering the time span from \numrange{2008}{2010} and \numrange{2011}{2018}. In total, we retrieved \num{6220} observations of {GJ~3473} from the archive. They were observed with telescopes \num{01} and \num{04} in the broad {RG715} filter. For the photometric analysis of the host star, we use the individual nightly binned time series, which shows a median internal uncertainty of \SI{2.6}{ppt} and a rms of \SI{8.7}{ppt} around the mean.

\subsection{High-resolution imaging}
As part of the standard process for validating transiting exoplanets and to assess the possible contamination of bound or unbound companions on the derived planetary radii \citep{Ciardi.2015}, high-resolution images of {GJ~3473} were taken within the TFOP `High Resolution Imaging' SG3 subgroup.

\paragraph{Gemini/NIRI.}
Nine images of {GJ~3473} in the Br$\gamma$ narrow filter $(\lambda_0 = 2.1686; \Delta\lambda = \SI{0.0295}{\mu\meter}$) were taken with the NIRI instrument mounted at the 8.1\,m Gemini North telescope \citep{Hodapp.2003} {on 22~March~2019} {as part of the Gemini programme GN-2019A-LP-101}. The science frames had an exposure time of \SI{3.5}{s} each and were dithered in a grid pattern with \SI{\sim 100}{px} spacing (\SI{\sim 2.2}{arcsec}). A sky background image was created by median combining the dithered images. The basic reduction included bad pixel interpolation, flatfield correction, sky background subtraction, and alignment and co-adding of the images. 

\paragraph{Keck/NIRC2.}
The Keck Observatory observations were made with the NIRC2 instrument on {the} 10.0\,m Keck II {telescope} behind the natural guide star AO system \citep[][and references therein]{Service.2016}.
The observations were taken on 25~March~2019 in the standard three-point dither pattern that is used with NIRC2 to avoid the left lower quadrant of the detector, which is typically noisier than the other three quadrants. The dither pattern step size was \SI{3}{arcsec} and was repeated four times.  
The observations were also obtained in the narrow-band Br$\gamma$ filter $(\lambda_0 = 2.1686; \Delta\lambda = \SI{0.0326}{\mu\meter}$) with an integration time of \SI{20}{\second} with one coadd per frame for a total of \SI{300}{\second} on target.  The camera was in the narrow-angle mode with a full field of view of $\sim10$\,arcsec and a pixel scale of 0.099442\,arcsec\,pix$^{-1}$.%

\section{Properties of {GJ~3473}}
\label{sec:stellar_prop}

\begin{table}
\caption{Stellar parameters of {GJ~3473}}
\label{tab:stellar_parameters}
\centering
\begin{tabular}{l c r}
\hline\hline
Parameter & Value & Ref. \\
\hline
\noalign{\smallskip}
\multicolumn{3}{c}{\em Name and identifiers}\\
\noalign{\smallskip}
{Name} &  {GJ~3473} & {Gli91} \\
{Alt. name} & {G~50--16} & {Gic59} \\
Karmn & J08023+033 & Cab16\\
TIC & 452866790 & Stas19 \\
TOI & 488 & \emph{TESS} {releases} \\
\noalign{\smallskip}
\multicolumn{3}{c}{\em Coordinates and spectral type}\\
\noalign{\smallskip}
$\alpha$ (J2000) & 08\,02\,22.88 & \emph{Gaia} DR2\\
$\delta$ (J2000) & +03\,20\,19.7 & \emph{Gaia} DR2\\
Sp. type & M4.0\,V &Haw96 \\
$G$\,[mag]& $12.4650\pm0.0003$&\emph{Gaia} DR2\\
{$T$\,[mag]}& {$11.1972\pm0.0073$} & {Stas19} \\
$J$\,[mag]& $9.627\pm0.023$ & 2MASS\\
\noalign{\smallskip}
\multicolumn{3}{c}{\em Parallax and kinematics}\\
\noalign{\smallskip}
$\mu_{\alpha}\cos\delta$\,[\si{mas\,yr^{-1}}]& $-403.17\pm0.09$&\emph{Gaia} DR2 \\
$\mu_\delta$\,[\si{mas\,yr^{-1}}]& $-381.01\pm0.05$&\emph{Gaia} DR2\\
$\pi$\,[\si{mas}]&$36.52\pm0.05$ &\emph{Gaia} DR2 \\
$d$\,[\si{pc}]& $27.39\pm0.04$ &\emph{Gaia} DR2 \\
$\gamma$ [\si{\kilo\meter\per\second}] & $-1.101\pm0.011$ & This work \\
$U$\,[\si{\kilo\meter\per\second}]& $-3.11\pm0.05$ & This work \\
$V$\,[\si{\kilo\meter\per\second}]& $-27.66\pm0.06$ & This work\\
$W$\,[\si{\kilo\meter\per\second}]& $-66.44\pm0.07$ & This work \\
\noalign{\smallskip}
\multicolumn{3}{c}{\em Photospheric parameters}\\
\noalign{\smallskip}
$T_\tx{eff}$\,[K]& $3347\pm54$ & This work \\
$\log{g}$\,[dex]& $4.81\pm0.06$ & This work \\
$[\tx{Fe/H}]$\,[dex]& $+0.11\pm0.19$ & This work \\
\noalign{\smallskip}
\multicolumn{3}{c}{\em Physical parameters}\\
\noalign{\smallskip}
$L_\star$\,[{$L_\odot$}]& {$0.01500\pm0.00019$}& This work \\
$R_\star$\,[$R_\odot$]& $0.364\pm0.012$ & This work \\
$M_\star$\,[$M_\odot$]& $0.360\pm0.016$ & This work \\
\noalign{\smallskip}
\multicolumn{3}{c}{\em Activity parameters}\\
pEW (H$\alpha$) [\si{\angstrom}] & $+0.08\pm0.15$ & This work \\
$\log R'_\mathrm{HK}$\,{[dex]} & $-5.62\pm0.22$ & This work \\
$v\sin{i}$\,[\si{\kilo\meter\per\second}]& $< 2$ &   This work\\
$P_\mathrm{rot}$ [d] & $168.3\pm4.2$ & This work \\
\noalign{\smallskip}
\hline
\end{tabular}
\tablebib{
    {Gli91: \citet{Gliese.1991};}
    {Gic59: \citet{Giclas.1959};}
    Cab16: \citet{Caballero.2016};
    Stas19: \citet{Stassun.2019};
    \emph{Gaia} DR2: \citet{GaiaCollaboration.2018};
    Haw96: \citet{Hawley.1996};
    2MASS: \citet{Skrutskie.2006}.
}
\end{table}

The star {\object{GJ~3473}} (G~50--16,  Karmn~ J08023+033) was included in the \emph{TESS} Input Catalogue as TIC~452866790 and declared a \emph{TESS} Object of Interest (TOI) \num{488} after the transiting planet candidate was found. A summary of the stellar parameters is given in \autoref{tab:stellar_parameters}. 
The star was classified as an {M4.0~V} star by \cite{Hawley.1996}, but it was never subject to an in-depth analysis of its properties \citep[e.g.][]{Newton.2014}. To determine precise stellar parameters, we used the high-resolution spectra from the CARMENES follow-up observations. Following \cite{Passegger.2018, Passegger.2019} and applying a measured upper limit of $v \sin i = \SI{2}{\kilo\meter \per\second}$, we calculated the effective temperature $T_\mathrm{eff}$, $\log g$, and $[\mathrm{Fe/H}]$ from a fit to a grid of PHOENIX stellar atmosphere models \citep{Husser.2013} using a $\chi^2$ method.
{The derived $T_\mathrm{eff}$ matches the literature spectral type taking into account {the} uncertainties in both parameters \citep{AlonsoFloriano.2015, Passegger.2018, Cifuentes.2020}.}
Next, we determined the bolometric luminosity, $L_\star$, by integrating the spectral energy distribution in {14} broad passbands from optical $B$ to $W4$ with the {\em Gaia} DR2 parallax \citep{GaiaCollaboration.2018} as in \cite{Cifuentes.2020}.
The radius, $R_\star$, was subsequently calculated using the Stephan-Boltzmann law. Lastly, the mass, $M_\star$, was derived from the \cite{Schweitzer.2019} empirical mass-radius relation derived from eclipsing binaries. {The values determined in this way are consistent with the mass and radius {determined from} \texttt{isochrones} \citep{Morton.2015}.}   
We updated the Galactocentric space velocities $UVW$ according to \cite{CortesContreras.2016} by combining the \emph{Gaia} DR2 parameters and the absolute velocity measured from the cross-correlation function (CCF) of the spectra with a weighted binary mask \citep{Lafarga.2020}. Using the space velocities and the \texttt{BANYAN}~$\Sigma$ Bayesian classification tool, we found no indication that {GJ~3473} is a member of any nearby young stellar association. {Instead, it is} most probably a field star located in the Galactic thin disk \citep{Gagne.2018}.

{GJ~3473} is listed in the Washington Double Star catalogue \citep{Mason.2001} as the primary of the binary system LDS\,5160 (Luyten Double Star), with a date of first satisfactory observation in 1949.
The secondary, at an angular separation of 49.29$\pm$0.09\,arcsec to the southeast of {GJ~3473} ($\theta$ = 222.8$\pm$14.1\,deg), is \object{LP~544--12} (GJ~3474, source \#3 in \autoref{fig:tpf_plot}), a $J$ = 12.2\,mag M6\,V star that shares the proper motion and parallax values of our planet-host star, but with a third of its mass \citep{Luyten.1979, Reid.1995, Newton.2017}.
At the distance of {GJ~3473}, the angular separation between the two stars translates into a projected physical separation of {1349.9$\pm$3.1\,au}.

\section{Analysis and results}
\label{sect:analysis}

\subsection{Transit search within the \emph{TESS} light curve}
\label{subsec:transit_search}

A transiting planet candidate around {GJ~3473} was announced on 14 March 2020 via the {\emph{TESS}} releases website\footnote{\url{https://tess.mit.edu/toi-releases}}. The candidate passed all tests from the SPOC Data Validation Report {\citep{Twicken.2018, Jenkins.2002, Li.2019}} and it is listed on the Exoplanet Follow-up Observing Program (ExoFOP)\footnote{\url{https://exofop.ipac.caltech.edu/tess/target.php?id=452866790}} webpage as having a period of \SI{1.1981}{d} and a transit depth of \SI{1.051}{ppt}.
We performed an independent transit search on the PDC-SAP light curve using the Transit-Least-Squares method \citep[TLS;][]{Hippke.2019}\footnote{\url{https://github.com/hippke/tls}}.
{We consider a signal to be significant if it reaches a signal detection efficiency \citep[SDE;][]{Alcock.2000, Pope.2016} of at least $\mathrm{SDE} \geq 8$.} The TLS shows a highly significant transit signal ($P \approx \SI{1.1979}{\day}$) with a{n} {SDE} of $\sim 18.4$  and a transit depth of \SI{1.071}{ppt}. After pre-whitening of the photometric data by fitting for this signal, a TLS of the residuals show{s} no {remaining} significant signals with $\mathrm{SDE} \geq 8$.

\subsection{{Adaptive-optics} imaging and limits of photometric contamination}
\label{subsec:hci}

\begin{figure*}
\centering
\includegraphics[]{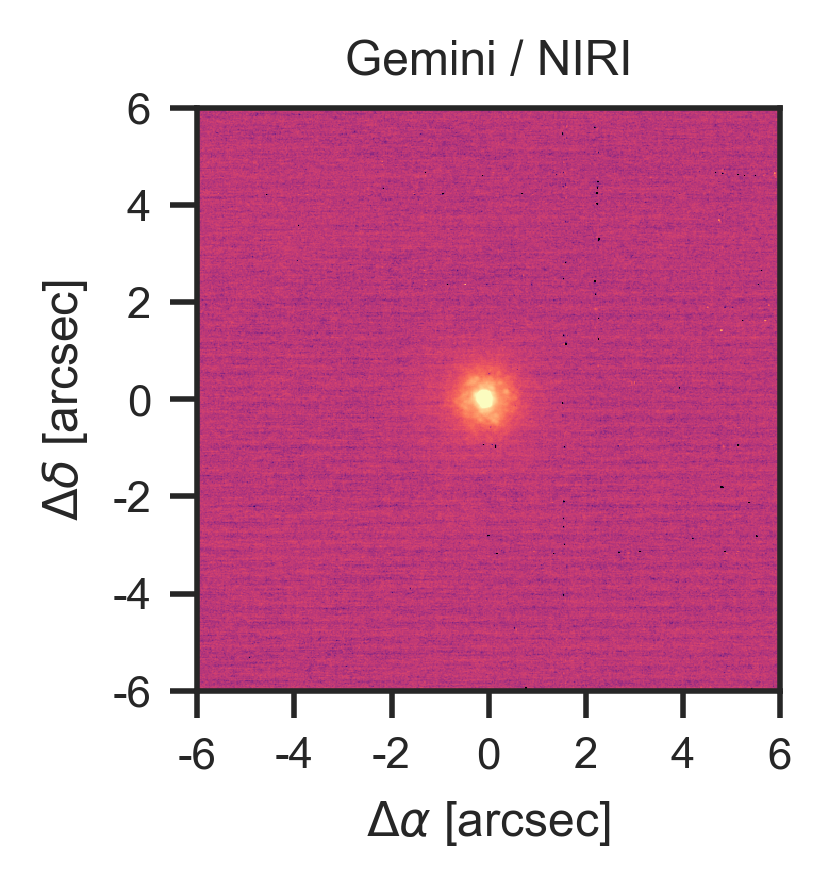}
\includegraphics[]{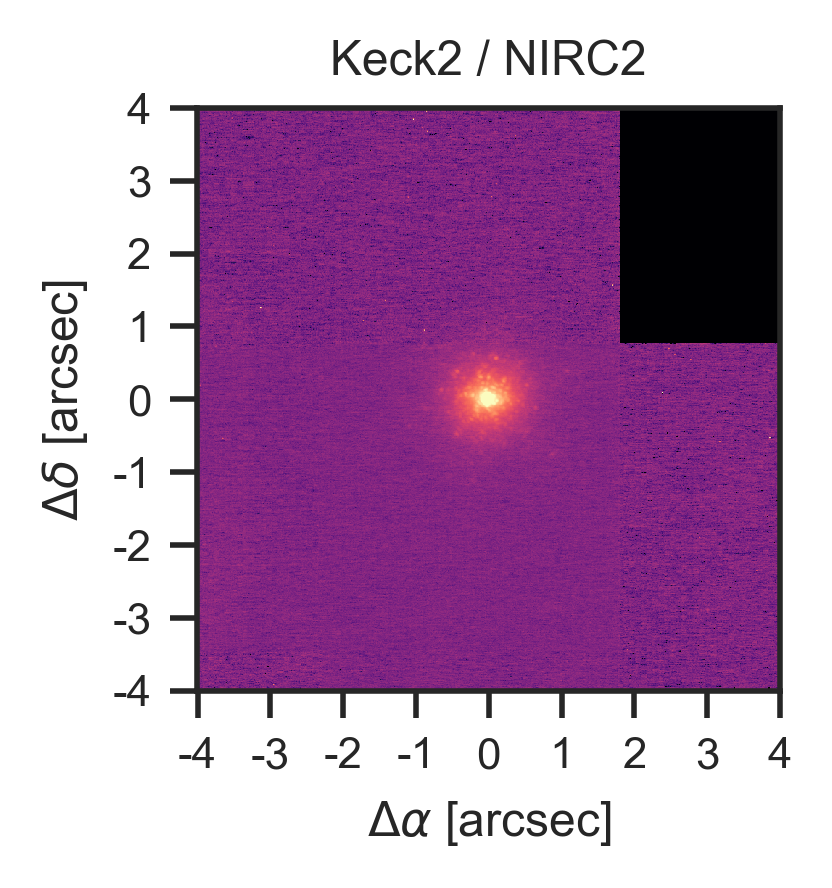}
\includegraphics[]{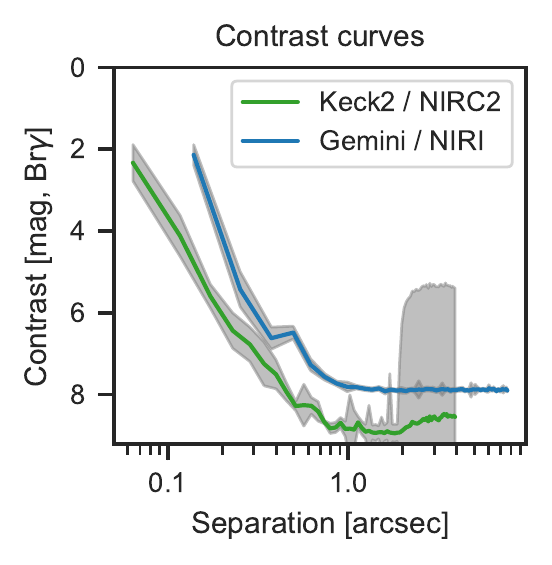}
\caption{AO images and contrast curves of the Keck II and Gemini North observations of {GJ~3473}.
The grey shaded regions in the contrast curves are the uncertainty, which apparently {rises dramatically} for NIRC2 because of a dead quadrant.}
\label{fig:hci}
\end{figure*}

As part of our standard process for validating transiting exoplanets and to assess the possible contamination of bound or unbound companions on the derived planetary radii \citep{Ciardi.2015}, we investigated the deep AO images from NIRC2 at Keck II and NIRI at Gemini North shown in \autoref{fig:hci}. {Both images were taken in Br$\gamma$ narrow filters.} 
No companions are visible to a separation of \SI{7.5}{arcsec}. The contrast curves are obtained by injecting fake sources of different brightness at different separations from the star and determining the radial $5\sigma$ detection limit. The NIRC2 observations are sensitive up to a contrast of $\Delta m =\SI{8.3}{mag}$ at a separation of \SI{0.5}{arcsec} to the star, but only span the region of the inner \SIrange{3}{4}{arcsec}. The area further out up to \SI{6.0}{arcsec} is covered by the NIRI image, which reaches a contrast level of $\Delta m = \SI{7.9}{mag}$. 
We therefore conclude, based on a combination of the contrast curves, a visual inspection of the AO images in \autoref{fig:hci}, and the \citet{Baraffe.2003} COND models for an approximate{} solar age, that {GJ~3473} does not have {a high-mass brown dwarf or more massive companion} at 13--160\,au.
Furthermore, using {additional} 2MASS imaging and {\em Gaia} DR2 astro-photometry, we ruled out the presence of stellar companions of any mass at separations beyond 160\,au up to a few thousands au (excluding the known companion LP~544--{12}).
{Another indicator for binarity is the re-normalised a posteriori mean
error of unit weight (RUWE), which quantifies the goodness-of-fit of the astrometric solution in the \emph{Gaia} DR2 \citep{Arenou.2018, Lindegren.2018}.} {At approximate separations between 1.3\,au and 13\,au}, any hypothetical stellar companion would {cause} {GJ~3473} to have a {\em Gaia} RUWE value larger than 1.41 \citep[its actual value is 1.06;][and references therein]{Cifuentes.2020} and an asymmetric point spread function in the NIRC2 and NIRI images.
{At separations closer than 1.3\,au, we would see a double-peak{ed} CCF or a long-term trend in the CARMENES RV data}.
We cannot exclude, however, the presence of substellar objects of a few Jupiter masses at wide separations (with orbital periods much longer than the RV coverage) {or {unfavourably} aligned objects at close separations}.

Additionally, we assessed the photometric contamination of the {\em TESS} light curves using Eq.~6 from {\cite{Espinoza.2019}}. From the AO images, we obtained upper limits from \SIrange{5}{8}{mag} in contrast for the inner area from  \SIrange{0.15}{7.5}{arcsec}, which correspond to contamination below \SI{1}{\percent}. For the nearby \emph{Gaia} sources apparent in \autoref{fig:tpf_plot}, we make use of the fact that the \emph{TESS} and \emph{Gaia} $G_{RP}$-band filter are very similar. We find that for the brightest nearby source (\# 3 in \autoref{fig:tpf_plot}), which is its binary companion LP~544--12, the dilution factor would be 0.96. However, given the separation of \SI{48.9}{arcsec} to {GJ~3473}, {this} is negligible and, thus, we assume for our modelling that there are no contaminating sources nearby.

\subsection{Transits only modelling}
\label{subsec:transit_only}
In order to refine the orbital period of the transiting planet candidate that was determined from the TLS analysis and to evaluate whether the individual follow-up observations show adequate transit detections, we first investigated the photometric observations separately from the RV measurements.\par
For all modelling tasks in this work, we {used} \texttt{juliet}\footnote{\url{https://juliet.readthedocs.io/en/latest/}} \citep{Espinoza.2019},
a fitting tool that uses nested sampling to efficiently evaluate the parameter space of a given prior volume and to allow for model comparison based on Bayesian evidences. {Here, \texttt{juliet} combines publicly available packages for RVs and transits, namely,
\texttt{radvel}\footnote{\url{https://radvel.readthedocs.io/en/latest}} \citep{Fulton.2018} and \texttt{batman}\footnote{\url{https://www.cfa.harvard.edu/~lkreidberg/batman/}} \citep{Kreidberg.2015}.} 
It allows us to choose among a range of different nested sampling algorithms for the fitting.
We opted for \texttt{dynesty}\footnote{\url{https://github.com/joshspeagle/dynesty}} \citep{Speagle.2020} because of its simple usage with regard to multi-processing. Additionally, \texttt{juliet} provides the implementation of Gaussian processes in the models using either \texttt{george}\footnote{\url{https://george.readthedocs.io/en/latest/}} \citep{Ambikasaran.2015} or \texttt{celerite}\footnote{\url{https://celerite.readthedocs.io/en/stable}} \citep{ForemanMackey.2017}.

As a first step, we modelled all of the \num{15} available ground based follow-up observations of transit events obtained by the TFOP SG1\footnote{As of 13 March 2020} separately with the \emph{TESS} light curve while fitting for the transit centre of each transit \citep[see][and the documentation of \texttt{juliet} for details of the implementation]{Eastman.2019}. In doing so, we {re-parametrised} the scaled {semi-major} axis to the stellar density, $\rho_*$. In this manner we can make use of the derived stellar parameters to obtain a density estimation as a fit prior. Furthermore, we implement the parameter transformation suggested by \cite{Espinoza.2018} and fit for the parameters, $r_1$ and $r_2,$ instead of the planet-to-star radius ratio, {$p$,} and the impact parameter, {$b$}. A quadratic limb-darkening model is used for the \emph{TESS} data \citep{Espinoza.2015}, which is parametrised by the $q_1$ and $q_2$ parameters \citep{Kipping.2013}, while a linear model is used for the ground-based follow-up observations. We used a linear term to detrend the LCOGT and MuSCAT light curves with airmass, while the MuSCAT2 light curves were pre-detrended (see \autoref{subsec:data_transit}). Based on the results from \autoref{subsec:hci}, we fix the dilution factor to one for all instruments, but consider free individual instrumental offsets. Also instrumental jitter terms are taken into account and added in quadrature to the nominal instrumental errorbar.

By carrying out this pre-analysis, we were able to disregard observations that show no, or only marginal transits, or seem to be only apparent transits with transit centres far from a linear ephemeris. The final dataset, which is presented in \autoref{sec:data} and which we use for the analysis in this work, includes \num{7} transit events with \num{11} observations of firm transit detections (cf. \autoref{tab:phot_follwup}).

In the next step, we combine these observations and {then} fit for a common period and time of transit centre that serve as a basis for the joint analysis. In doing so, we determine $P = 1.1980034^{+0.0000022}_{-0.0000023}$\,d and $t_0 = 2458492.20410^{+0.00052}_{-0.00050}$.

\subsection{RV only modelling}
\label{subsec:rv_only}

\begin{figure}
\centering
\includegraphics[]{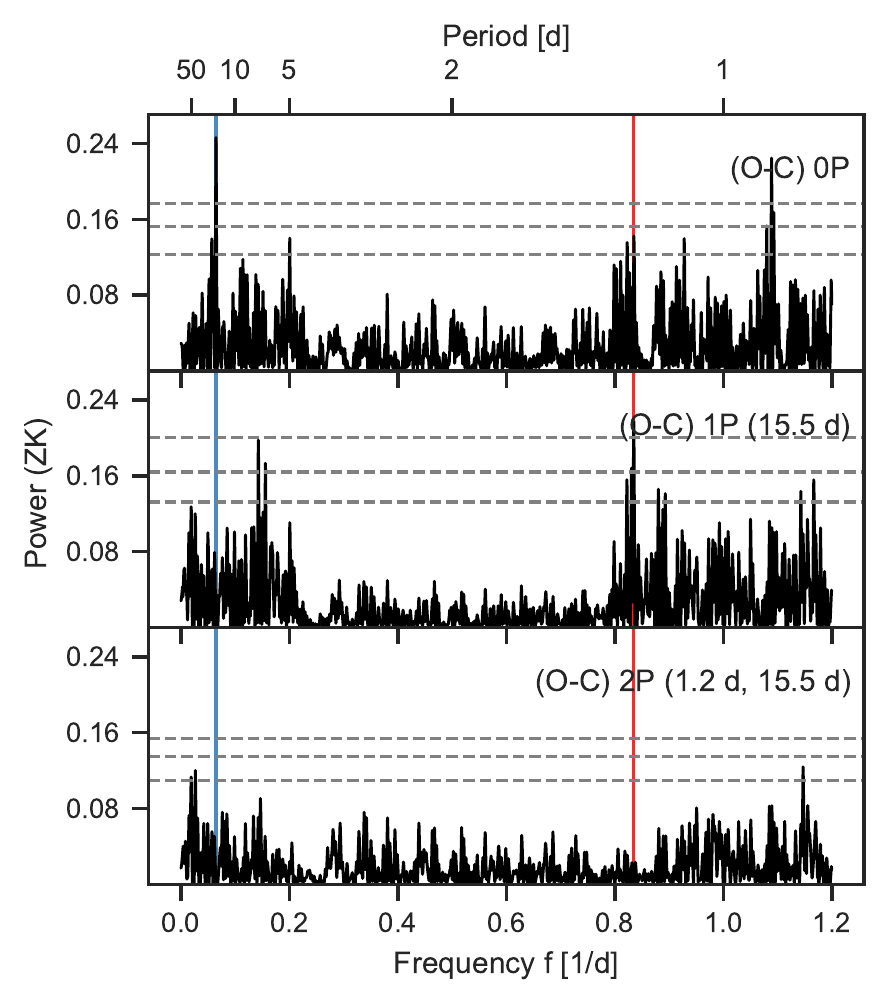}
\caption{GLS periodograms of the RV measurements.
Vertical lines mark the transiting planet (b, {solid red}) and the new RV planet (c, solid blue).
The horizontal dashed grey lines show the false alarm probability (FAP) of \SI{10}{\percent}, \SI{1}{\percent,} and \SI{0.1}{\percent} determined from \num{10000} random {realisations} of the measurements.}
\label{fig:periodogram_rv}
\end{figure}

We approach the analysis of the RVs with a signal search in the data, {proceeding} as if we do not know of the transiting planet a priori. In \autoref{fig:periodogram_rv}, the {generalised Lomb-Scargle periodograms \citep[GLS;][]{Zechmeister.2009}} of the residuals from different fits of increasing complexity to the combined RVs from CARMENES, IRD, and HARPS are shown. {We normalised the periodograms using the parametrisation of \citet[ZK]{Zechmeister.2009}. For all fits, we used Gaussian distributed priors for the signal of the transiting planet candidate based on the results from \autoref{subsec:transit_only} and a narrow uniform range around the peak of the second signal. Instrumental offsets and jitter are treated separately for each dataset. For comparison, we {list} the Bayesian evidences of the fits in \autoref{tab:rv_fit_evidence}.}

The first panel in \autoref{fig:periodogram_rv} shows the periodogram of the residuals after fitting a flat model that only includes offsets and instrumental jitter to the CARMENES, IRD, and HARPS measurements. The strongest signal apparent in the RV data is a period at \SI{15.5}{d}. After subtracting this periodicity with a circular Keplerian fit, the residual periodogram shows {a significant signal ($\mathrm{FAP} < \SI{1}{\percent}$),  coincident with the period of the transiting planet at $P \approx \SI{1.198}{\day}$ (see the middle panel of \autoref{fig:periodogram_rv}).} The FAP for a signal to occur especially at this frequency can be evaluated using the method by \cite{Baluev.2008} and the power of the signal in a Lomb-Scargle periodogram. By this means, we find a spectral $\mathrm{FAP} \approx \SI{0.003}{\percent}$. We verify this using a bootstrap method of \num{1e6} random {realisations} over a decreasing frequency range {centred} on the period in question, which yields $\mathrm{FAP}\approx \SI{0.002}{\percent}$. {This is in agreement with the Baluev method and we therefore assume a FAP of \SIrange{0.002}{0.003}{\percent} for the signal to occur at the expected period. Furthermore, the phase of this signal matches the phase of the planet candidate from \emph{TESS} and we thus report a highly significant detection of the transiting planet candidate in the RV.}
The two other signals of significant power at periods of \SI{6.41}{\day} and \SI{7.00}{\day} are aliases of the former signal of the transiting planet due to the approximately daily sampling. This is reflected by the fact that they {disappear} when the \SI{15.5}{d} signal is fitted together with the period of the transiting planet at $P \approx \SI{1.198}{\day}$ (see the bottom panel of \autoref{fig:periodogram_rv}). Given that RV data provide more information on the eccentricity of an orbit, we performed an analogous analysis using eccentric orbits. We find that the difference between a circular and eccentric orbit is indistinguishable ($\Delta\ln{\mathcal{Z}} = -0.45$) and, therefore, we use the results for the simpler circular model fits. The residuals of this fit comprising two circular Keplerian signals do not show any further periodicities with FAPs {above} our significance criterion of \SI{1}{\percent}.

\subsection{Joint modelling}
\begin{figure*}
\centering
\includegraphics[]{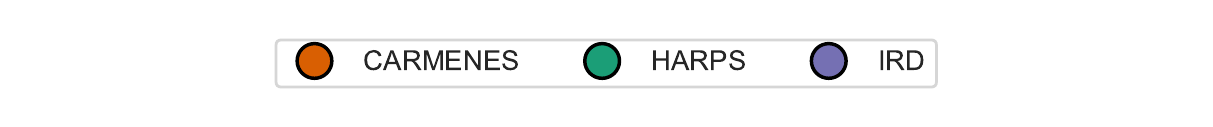}

\includegraphics[]{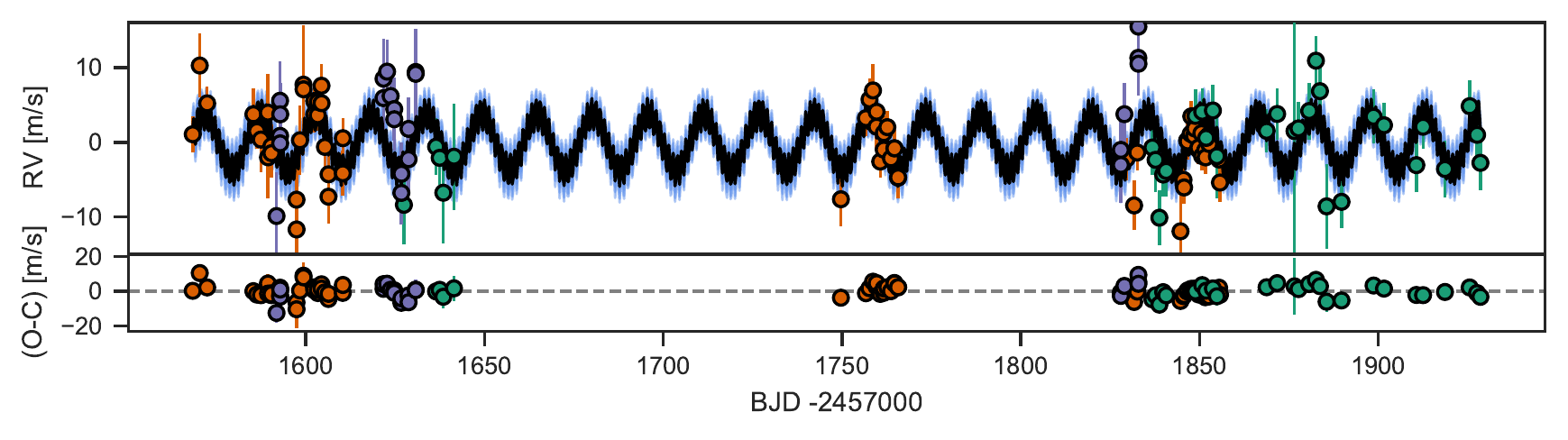}

\includegraphics[]{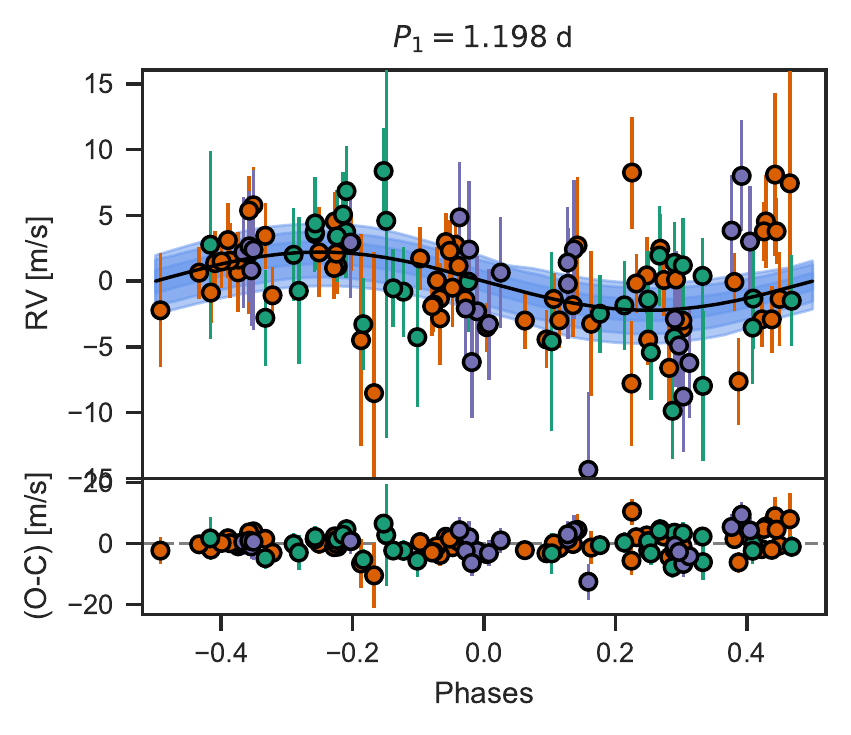}
\includegraphics[]{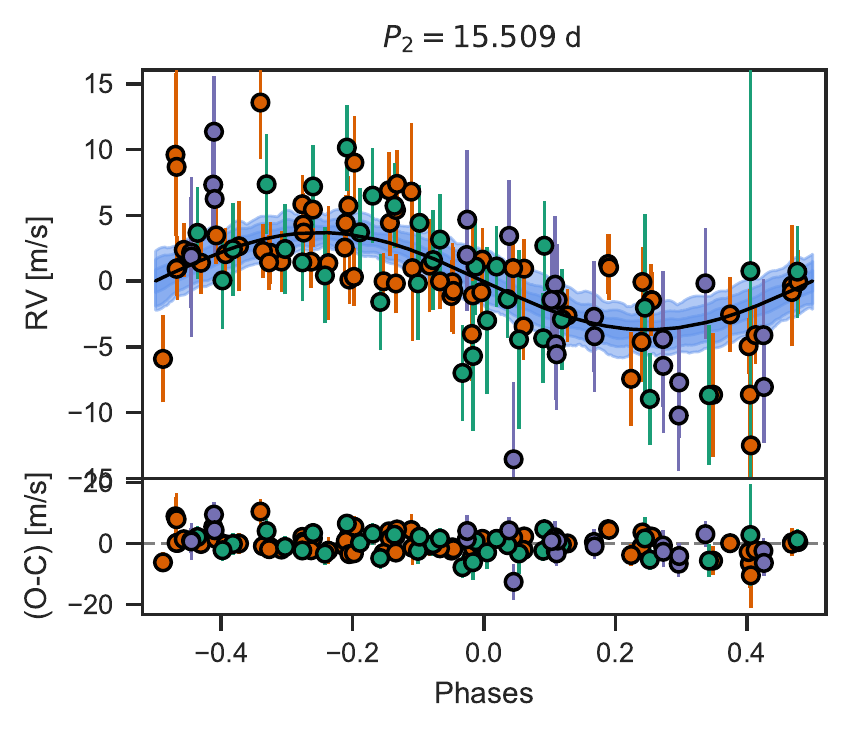}
\caption{Results from the joint fit for the RV data of  CARMENES, IRD, and HARPS. The top part of each panel shows the measurements as coloured circles -- errorbars include the instrumental jitters added in quadrature -- and the median of the best-fit \texttt{juliet} model as the black curve. The grey shaded areas mark the \SI{68}{\percent}, {\SI{95}{\percent,} and \SI{99}{\percent} credibility intervals}. {To avoid overcrowding of the figure, we binned the IRD data, which were taken with a high cadence, to chunks of \SI{30}{\minute} each.} In the lower part, the residuals after the model is subtracted (O-C) are shown. \textit{Top panel:} RVs over time. \textit{Bottom panels:} RVs phase-folded to the periods of the transiting planet (\textit{left}) and the new RV planet (\textit{right}).} 
\label{fig:joint_fit_rv}
\end{figure*}

\begin{figure*}
\centering
\includegraphics[]{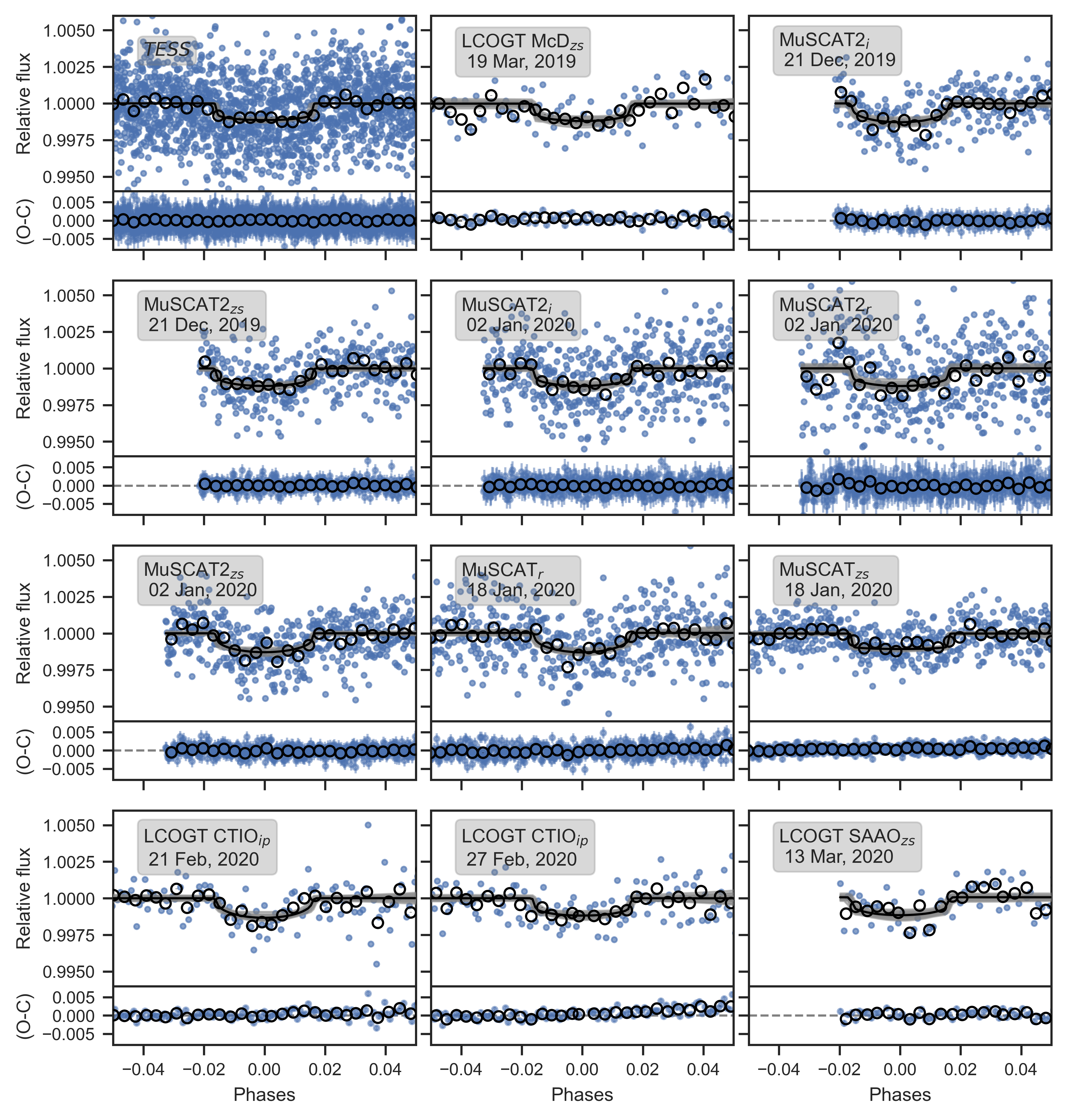}
\caption{Results from the joint fit for the transit observations. In the top part of each panel, the black curve presents the best-fit \texttt{juliet} model together with the \SI{68}{\percent}, {\SI{95}{\percent,} and \SI{99}{\percent} credibility intervals displayed by the grey shaded regions}. The observations of the respective instruments are phase-folded to the period of the transiting planet. For the fit, the individual data points (blue) are used, but the binned data are also shown for clarity (white circles). Error bars of the individual measurements with the instrumental jitter terms added in quadrature are only displayed in the bottom part of the panels, which show the residuals after subtracting the model (O-C). The names of the instruments and the dates of the observations are denoted in the grey boxes in the upper left corner of each panel.}
\label{fig:joint_fit_transit}
\end{figure*}
For the final retrieval of the most precise parameters we perform a joint fit of the \emph{TESS} observations, the ground-based transit follow-ups, and the combined RV data. The model includes two circular planets: firstly, the transiting planet that is detected in the photometry and RV data and; secondly, a non-transiting planet that is only apparent in the RV data. {The model of the joint fit comprises 58 free parameters, which turns the fit into a high-dimensional problem. A fit with uninformed priors would therefore be very costly. Hence, we} make use of the findings from the photometry-only analysis in \autoref{subsec:transit_only} and the RV-only analysis in \autoref{subsec:rv_only} -- that is, we implement Gaussian distributed priors for the planetary parameters, {as, for example, in \cite{Brahm.2019, Espinoza.2019b, Kossakowski.2019, Luque.2019}; or \cite{Bluhm.2020}. Since we use uninformed priors for the planetary parameters for the transit and RV-only fits, nested sampling warrants an efficient exploration of the possible parameter space fitting the individual datasets. Planetary parameters specific to a given data type, such as the planetary semi-amplitude in RV data or planet-to-star radius (and others) in transit data, would not change significantly in a joint fit as they are independent from the other data to first order. Generally, this also holds  true for the shared parameters since they are mostly driven by either one or the other method. Likewise, using the posterior results from the transit-only and RV-only fits as a prior knowledge for the joint fit is therefore justified given that the chosen prior volume for the joint fit does not restrict the posterior volume from the individual fits. Following this, the width of the priors that we choose for the planetary parameters of the joint fit are three times the standard deviation of the posterior results from the individual best fits. It limits the computational cost, but still allows the nested sampling algorithm to freely explore the parameter space since the Gaussian distribution has no strict borders. In the end, the posterior distribution of our joint fit is much narrower than that of the input priors, which shows that the input priors were conservatively chosen to map the relevant parameter space and did not reject crucial possible solutions.} Descriptions and justifications of the {adopted} instrumental parameters and priors can be found in the respective subsections, \autoref{subsec:transit_only} and \autoref{subsec:rv_only}. A summary of the used priors is given in \autoref{tab:priors_joint_fit}.

In \autoref{fig:joint_fit_transit} and \autoref{fig:joint_fit_rv}, we show the final models of the joint fit based on the posterior of the sampling. The median posteriors {of the planetary parameters} are shown in \autoref{tab:posterior_planets} and the full list of the posteriors {of the instrumental parameters} is given in \autoref{tab:posterior_instruments}. 

{\setlength{\extrarowheight}{4pt}%
\begin{table}
    \centering
    \caption{Posterior parameters of the joint fit of the transit and RV data.}
    \label{tab:posterior_planets}
    \begin{tabular}{l c l} 
        \hline
        \hline
        Parameter & Posterior\tablefootmark{(a)} & Units \\
        \hline
        \noalign{\smallskip}
        \multicolumn{3}{c}{\textit{Stellar parameters}} \\
        \noalign{\smallskip}
        $\rho_\star$ & $10.93^{+0.66}_{-0.69}$ & \si{\gram\per\centi\meter\cubed} \\
        \noalign{\smallskip}
        \multicolumn{3}{c}{\textit{Planetary parameters}} \\
        \noalign{\smallskip}
        $P_b$                    & $1.1980035^{+0.0000018}_{-0.0000019}$      & d               \\
        $t_{0,b}$                & $2458492.20408^{+0.00043}_{-0.00042}$      & d               \\
        $r_{1,b}$                & $0.557^{+0.044}_{-0.049}$                  & \dots           \\
        $r_{2,b}$                & $0.03184^{+0.00069}_{-0.00067}$            & \dots           \\
        $K_{b}$                  & $2.21^{+0.35}_{-0.35}$                     & $\mathrm{m\,s^{-1}}$ \\
        $\sqrt{e_{b}}\sin \omega_{b}$    & $0$ (fixed)                        & \dots   \\
        $\sqrt{e_{b}}\cos \omega_{b}$    & $0$ (fixed)                        & \dots   \\
        \noalign{\smallskip\smallskip}
        $P_c$                    & $15.509^{+0.033}_{-0.033}$                 & d               \\
        $t_{0,c}$                & $2458575.62^{+0.42}_{-0.43}$               & d               \\
        $K_{c}$                  & $3.75^{+0.45}_{-0.42}$                     & $\mathrm{m\,s^{-1}}$ \\
        $\sqrt{e_{c}}\sin \omega_{c}$    & $0$ (fixed)                        & \dots   \\
        $\sqrt{e_{c}}\cos \omega_{c}$    & $0$ (fixed)                        & \dots       \\
        \hline
    \end{tabular}
    \tablefoot{
      \tablefoottext{a}{Error bars denote the $68\%$ posterior credibility intervals.
       The posteriors of the instrumental parameters are continued in \autoref{tab:posterior_instruments}.}
      }    
\end{table}}

\subsection{Stellar activity}
\label{subsec:activity}
\begin{figure}
\centering
\includegraphics[]{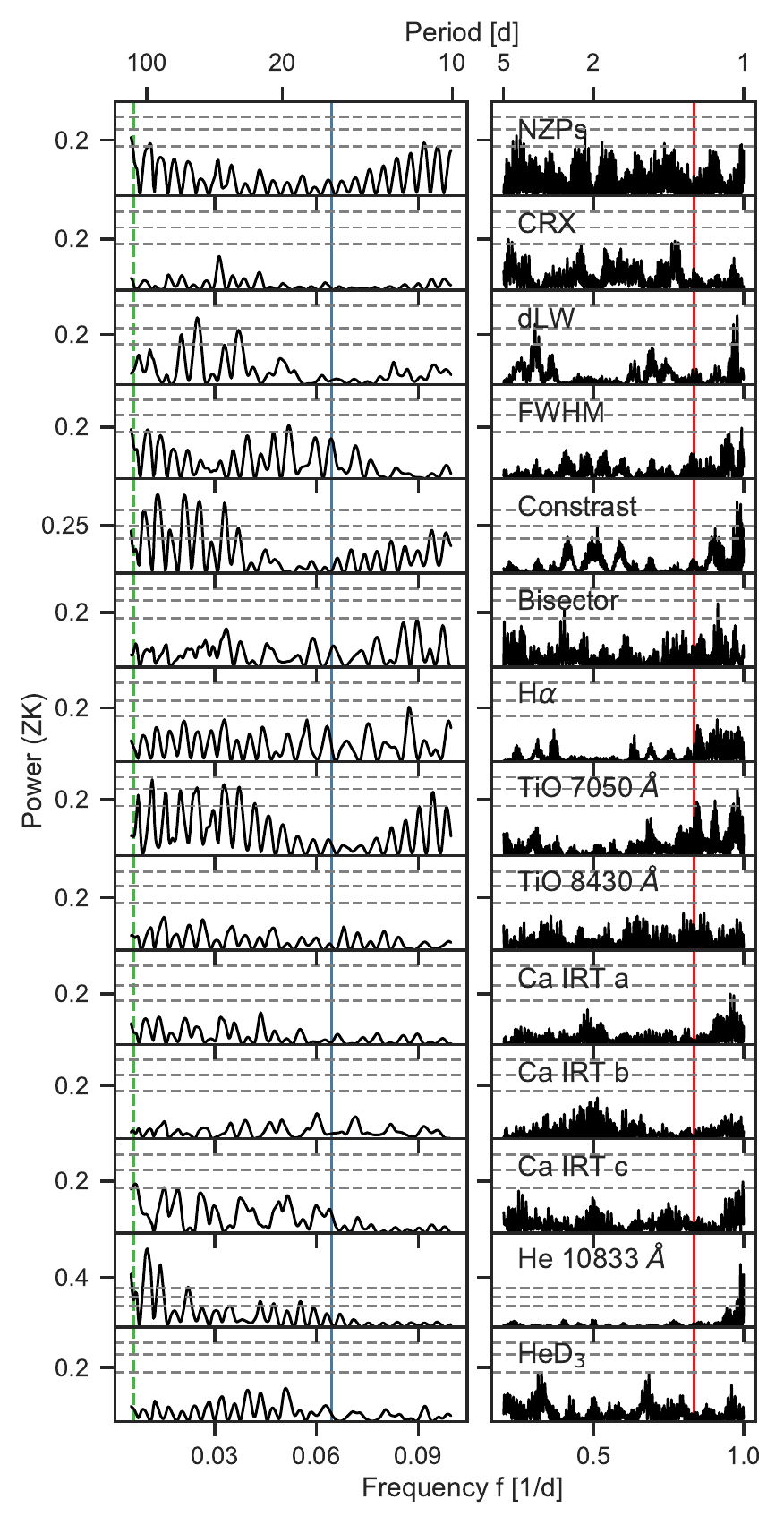}
\caption{GLS periodograms of a number of activity indicators based on spectroscopic data obtained by CARMENES, split into two frequency ranges.
The vertical lines mark the frequencies of the transiting planet candidate (red solid), the {\SI{15.5}{d}} periodicity visible in the RV (blue solid; see \autoref{subsec:rv_only}), and the determined photometric rotation period (dashed green; see \autoref{subsect:stellar_photometry}).
The horizontal grey lines show the false alarm probability (FAP) of \SI{10}{\percent}, \SI{1}{\percent,} and \SI{0.1}{\percent} determined from \num{10000} random {realisations} of the measurements.}
\label{fig:periodogram_act}
\end{figure}

We investigated a set of activity indicators derived from the CARMENES spectra to search for signals of stellar activity that would interfere with the transiting planet candidate or provide information on the origin of the second periodicity that is visible in the RV data (see \autoref{subsec:rv_only}). 
In \autoref{fig:periodogram_act}, the GLS periodograms of 13 selected activity indicators, as well as our applied nightly zero-point offsets, are shown. The chromatic index (CRX) and the differential line width (dLW) are products of the \texttt{SERVAL} reduction pipeline 
\citep[][]{Zechmeister.2018}.
From the CCF (see \autoref{sec:stellar_prop}), the full-width at half-maximum (FWHM), the contrast, and the bisector span are determined \citep{Lafarga.2020}. The pseudo-equivalent width after subtraction of an inactive template spectrum (pEW$'$) of the chromospheric H$\alpha$, Ca~{\sc ii}~IRT (a, b and c),  He~{\sc i}~$\lambda$\SI{10833}{\angstrom} and He~{\sc i}~D$_3$ lines, and the photospheric TiO~$\lambda$\SI{7050}{\angstrom} and TiO~$\lambda$\SI{8430}{\angstrom} indices are calculated following \citet{Schofer.2019}.

A measured median pEW of the H$\alpha$ line of \SI[retain-explicit-plus]{+0.08\pm0.15}{\angstrom} indicates that {GJ~3473} is a rather inactive star \citep{Jeffers.2018}. {We find a significant, although moderate,  
correlation between RV and the CRX and Na~{\sc i} D activity indices, however, the GLS periodograms from the} extensive set of activity indicators do not show any power at the frequencies of the transiting planet candidate or the \SI{15.5}{\day} signal. The dLW, CCF contrast, TiO~$\lambda$\SI{7050}{\angstrom}, and He~{\sc i}~$\lambda$\SI{10833}{\angstrom} show a forest of signals with $\SI{1}{\percent} < \mathrm{FAP} < \SI{10}{\percent}$ in the range of approximately \SIrange{30}{100}{d}. This is consistent with a lower limit of the stellar rotation period to be longer than $\sim \SI{9}{d}$ as determined from $v \sin i < \SI{2}{\kilo\meter \per\second}$, however, there is no common periodicity or conclusive pattern, which would hint {at} the rotation period of the star. The most significant signal, which is apparent in the He~{\sc i}~$\lambda$\SI{10833}{\angstrom} indicator, has a period of around \SI{100}{\day}. From the HARPS spectra we derive $\log R'_\mathrm{HK} = -5.62\pm0.22$, which is equivalent to a stellar rotation period of \SI{109(37)}{d} following the $R'_\mathrm{HK}$ vs. $P_\mathrm{rot}$ relationship of \cite{AstudilloDefru.2017b}. We also investigated GLS {periodograms} of the HARPS activity indicators derived by the {\texttt{DRS}} pipeline, but we do not find any significant periodicity and, therefore, we do not present them here.

\subsection{Photometric stellar rotational period}
\label{subsect:stellar_photometry}
\begin{figure*}
\centering
\includegraphics[width=0.48\textwidth]{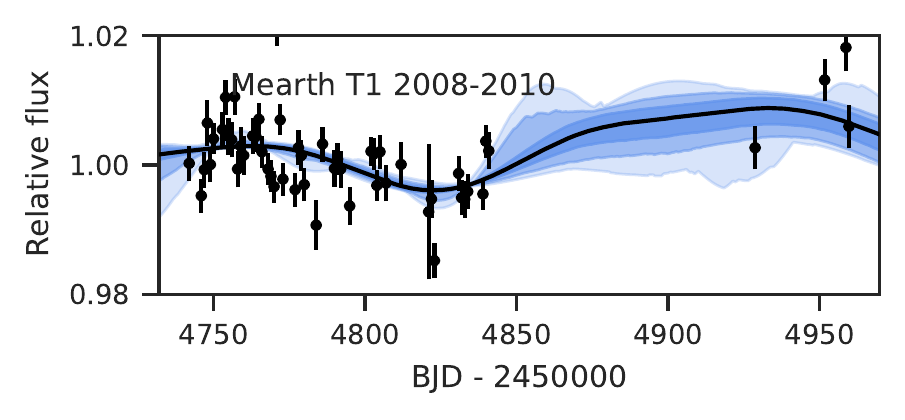}
\includegraphics[width=0.48\textwidth]{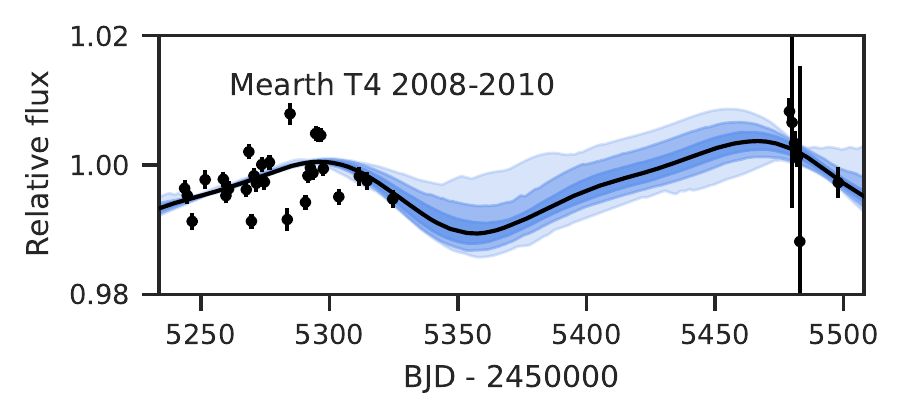} 

\includegraphics[width=0.48\textwidth]{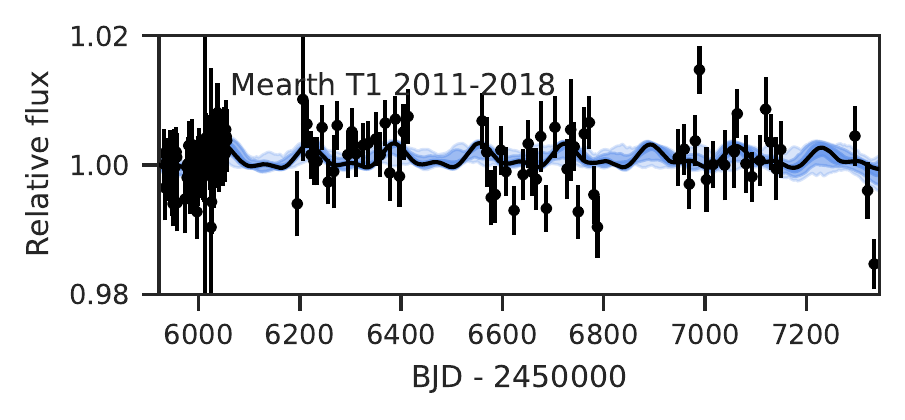}
\includegraphics[width=0.48\textwidth]{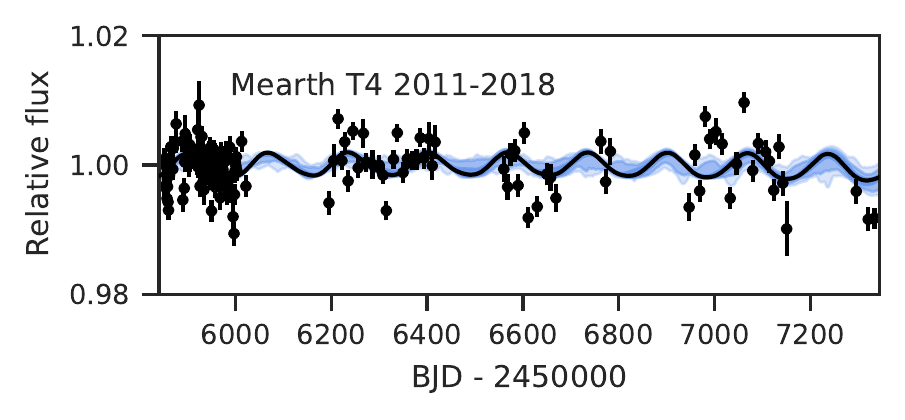} 

\includegraphics[width=0.48\textwidth]{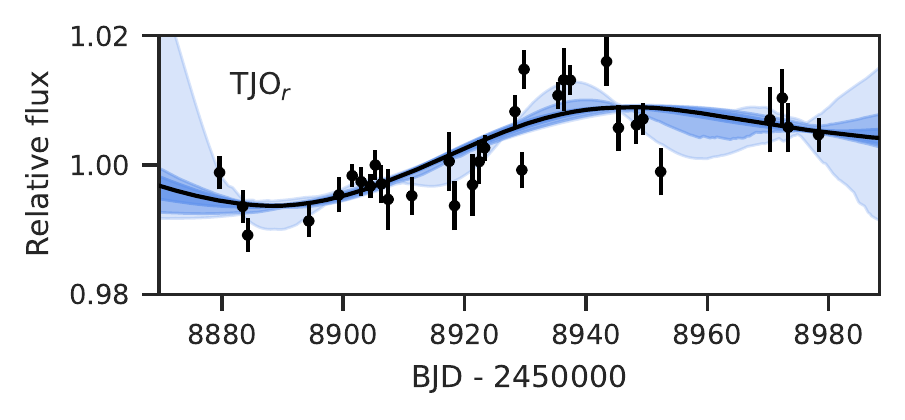}
\caption{Gaussian process fits to the photometric monitoring data of {GJ~3473}. From top to bottom: MEarth T1 2008-2010, MEarth T4 2008-2010, MEarth T1 2011-2018, MEarth T4 2011-2018, and TJO. {The black line shows the median GP model extracted for each instrument and the blue shades denote the \SI{68}{\percent}, \SI{95}{\percent} and {\SI{99}{\percent}} confidence intervals.}}
\label{fig: photometric_GP}
\end{figure*}

We combined the $R$-band TJO data collected between January and May 2020 and the RG715-band MEarth data taken between 2008 and 2018 to determine a stellar rotation period. 
A {marginalised} likelihood periodogram \citep[MLP;][]{Feng.2017} analysis of the combined data, where we fit for {jitter and offsets} between the datasets, indicated a preliminary periodicity of 160\,d.
The MLP uses sinusoidal functions to model possible significant signals. However, stellar activity tends to be quasi-periodic and can also deviate significantly from a simple sinusoidal. 
Thus, we used a Gaussian process (GP) to fit the photometry in a second approach.

We used \texttt{juliet} and select the quasi-periodic kernel by \texttt{george} for the modelling of the photometric data:
\begin{equation}
\label{Eq:GoergeGP}
k_{i,j}(\tau)=\sigma^2_{GP}\exp{(-\alpha\tau^2-\Gamma\sin^2{(\pi\tau/P_{\tx{rot}}})),}    
\end{equation}
where $\sigma_{GP}$ is the amplitude of the GP component given in {\si{ppt} (or \si{\meter\per\second} when applied to RV data)}, $\Gamma$ is the amplitude of GP sine-squared component, $\alpha$ is the inverse length-scale of the GP exponential component given in \si{\per\day\squared}, $P_{\tx{rot}}$ is the period of the GP quasi-periodic component given in days{,} and $\tau = |t_{i} - t_{j}|$ is the temporal distance between {two} measurements. {To perform a blind search for quasi-periodic signals with the GP model, we put in uninformed priors for $\sigma_{GP}$, $\Gamma_i,$ and $\alpha$, but take a uniform range from \SIrange{2}{200}{\day} for $P_{\tx{rot}}$.}

In doing so, the data of each instrument are {averaged into nightly bins} because of the large dataset and the computationally expensive log-likelihood evaluation of the used kernel. A daily sampling of the photometry is reasonable since we are searching for signals with periods of at least multiple days (see \autoref{subsec:activity}). Furthermore, binning reduces short-term variations due to jitter and decreases the uncertainties of the data points. For the GP model, we consider that each dataset can have different solutions for the amplitude parameters, $\sigma_{GP}$ and $\Gamma$. This accounts for the possibility that the stellar activity depends on wavelength and might impact each instrument differently. 
However, the timescale parameters, such as the rotational period, $P_{\tx{rot}}$, and the exponential decay of the signal $\alpha$, for example, due to spot-life time, should {not depend on the instrument}. 
For the latter two parameters, we therefore allow only for global solutions of the GP model. We also model the flux offset between the photometric datasets, as well as an extra jitter component, which is added in quadrature to the diagonal of the resulting covariance matrix. Our GP fit, using unconstrained priors (\autoref{tab:priors_rotation_period}), results in only one specific region within the prior volume that has a high density of posterior samples with high likelihood. 
We show the nightly binned photometric data and the GP fit with its uncertainties in Fig~\ref{fig: photometric_GP}.

From the posterior solutions we derive a photometric rotation period, $P_{\text{rot,phot}}=168.3^{+ 4.2}_{-3.1}$\,d for {GJ~3473}, which is consistent with the result from the MLP analysis and, within $2\sigma$, with the expected period from $\log{R'_{\rm HK}}$.
Both estimates show that {GJ~3473} is a slow rotator, which should not exhibit strong signals related to activity. {This is also in agreement with the spectroscopic activity indicators, which do not exhibit a predominant periodicity and no H$\alpha$ activity.}

\subsection{Investigation of the \SI{15.5}{d} signal}
\label{subsec:15dsignal}
\begin{figure*}
\centering
\includegraphics[]{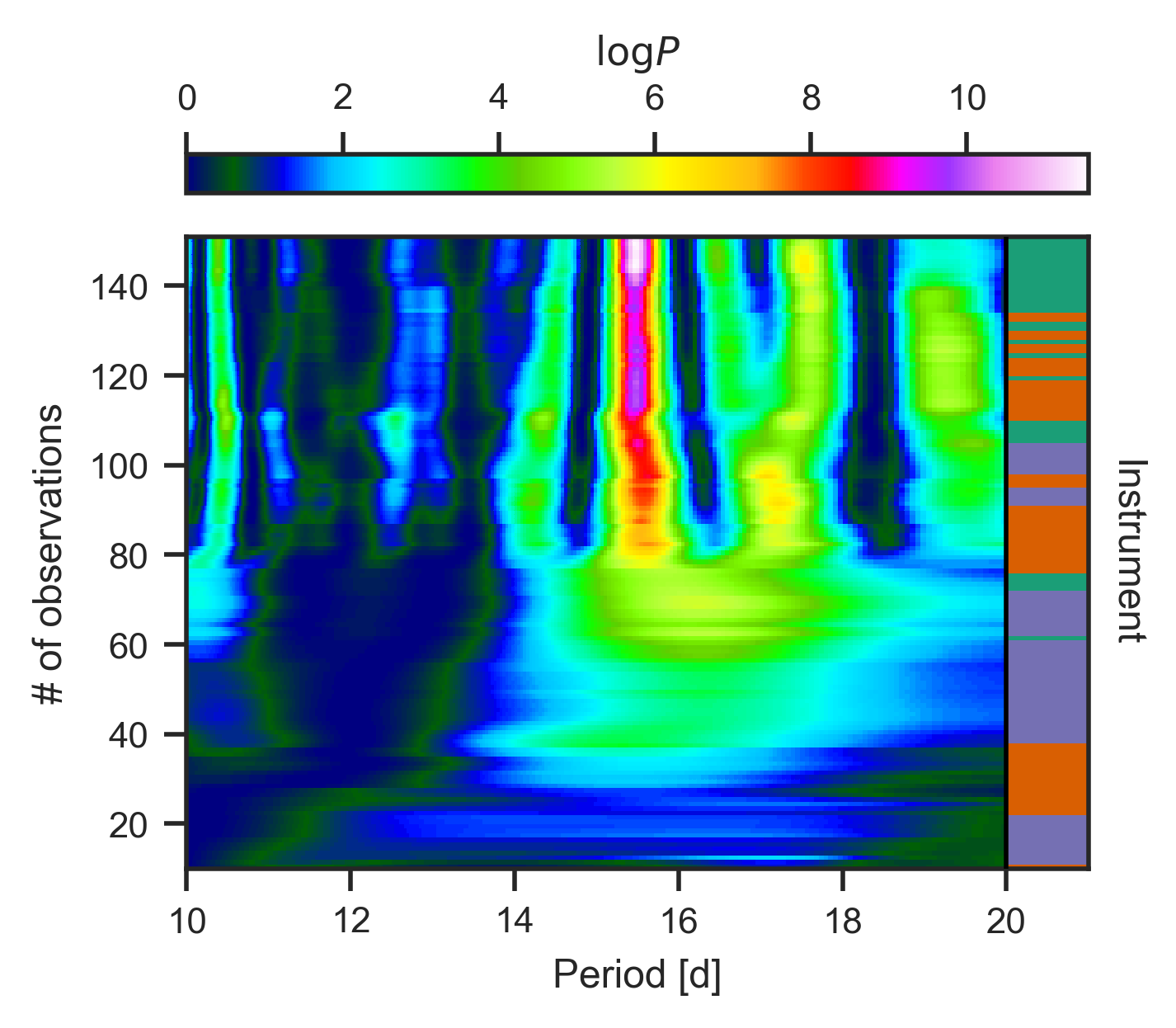}
\includegraphics[]{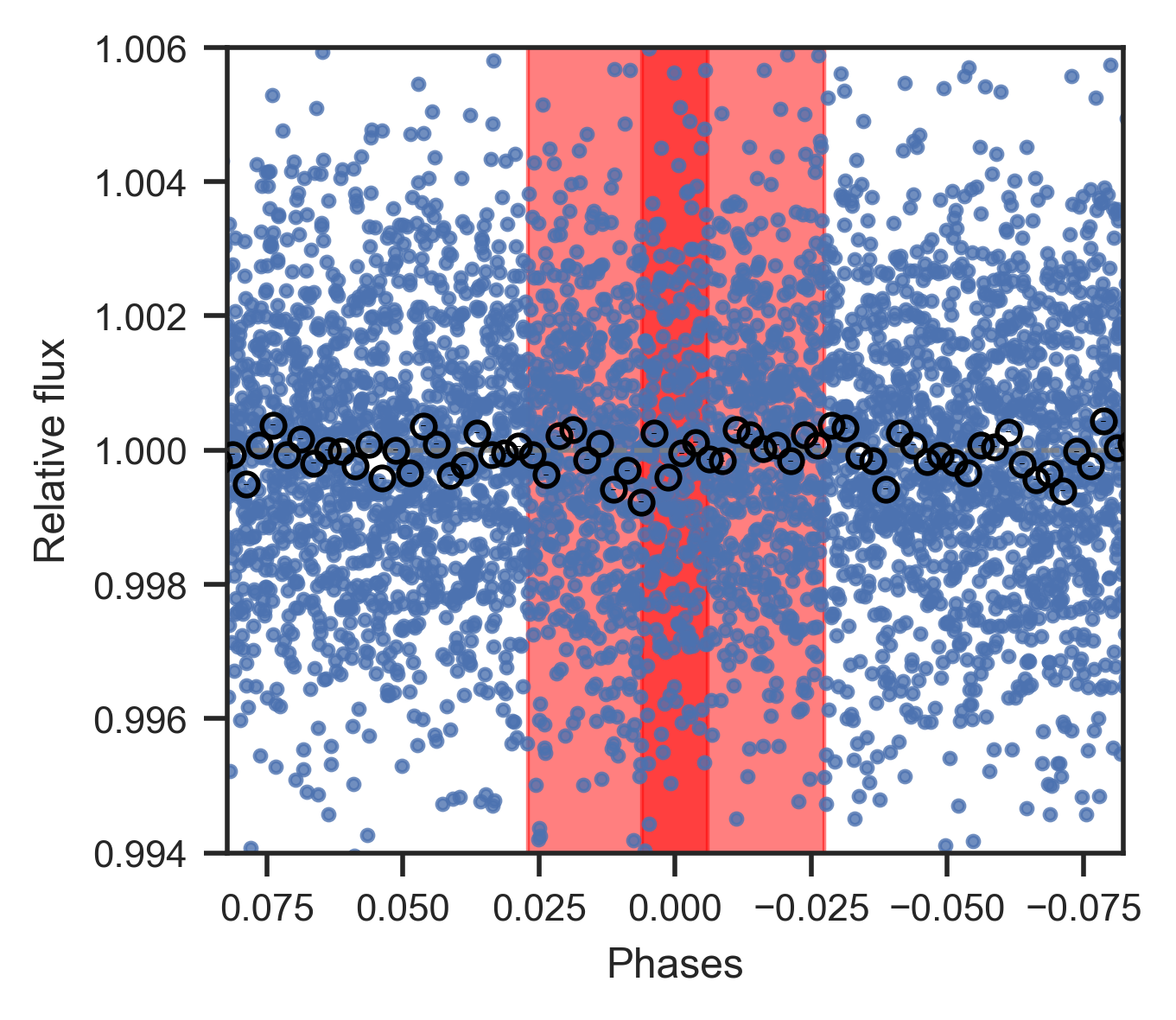}
\caption{Investigation of the \SI{15.5}{d} signal {\textit{Left:} Stacked BGLS periodogram of the residuals after fitting for the transiting planet. The colourbar on the right side indicates the instrument of the corresponding data point (orange: CARMENES, purple: IRD, green: HARPS). \textit{Right}: \emph{TESS} light curve {phase-folded} to the period and time of transit centre of {GJ~3473}~c as determined from the RVs. The saturated red shaded region indicates the expected transit, while the light red shaded region denotes the \SI{68}{\percent} credibility interval of the time of transit centre.}}
\label{fig:bgls15d}
\end{figure*}

\begin{table}
\caption{Bayesian log-evidences for the different models used to fit the RVs.}
\label{tab:rv_fit_evidence}
\centering
\begin{tabular}{l S c S}
\hline\hline
\noalign{\smallskip}
Model & {Periods\tablefootmark{(a)}} &$\ln{\mathcal{Z}}$ & {$\Delta\ln{\mathcal{Z}}$} \\
      & {[\si{\day}]}                        &                   &                          \\    
\noalign{\smallskip}
\hline
\noalign{\smallskip}
0 Planets & {\dots} & $-474.6\pm0.2$ & 0\\
1 Planet & 1.20 & $-468.1 \pm0.2$ &  6.5 \\
1 Planet &  15.5 & $-456.7\pm0.3$ & 17.9 \\
1 Planet + GP &  1.20  & $-444.9\pm0.3$ & 29.7 \\
2 Planets &  {15.5, 1.20} & $-442.7\pm0.3$ & 31.9 \\
\noalign{\smallskip}
\hline
\end{tabular}
\tablefoot{
        \tablefoottext{a}{Rounded to three digits.}
    }
\end{table}

The \SI{15.5}{d} signal seems to be unrelated {to} stellar activity or the stellar rotation period. Following \autoref{fig:joint_fit_rv}, the signal looks stable for the entire period of observations and shows no significant deviations from a circular Keplerian motion. {However, we thoroughly examined the signal in order to asses its nature and to test whether we can attribute it {unambigously} to a planetary origin.}

{We used \texttt{juliet} to perform a model comparison based on the Bayesian evidence of different models, applied to the RV data only, in order to check whether the \SI{15.5}{d} signal is indeed best fit with a Keplerian model. The log-evidences of the results are shown in \autoref{tab:rv_fit_evidence}. As outlined by \citet{Trotta.2008}, we consider a difference of $\Delta\ln{\mathcal{Z}} > 2$ as weak evidence that one of the models is preferred over the others and $\Delta\ln{\mathcal{Z}} > 5$ that a model is significantly favoured. We {use Gaussian distributed priors} based on the {posterior solutions} from \autoref{subsec:transit_only} {to account for} the transiting planet candidate{,} and {uniform priors} for instrumental offsets and jitter. However, we adopted two approaches to include the \SI{15.5}{d} signal in the modelling: on the one hand, a simple two-planet model is fitted to the data and on the other hand, we implement a quasi-periodic GP (see \autoref{Eq:GoergeGP} in \autoref{subsect:stellar_photometry}) to test the {possibility} that the second signal {does not have} a Keplerian nature and is only of a quasi-periodic {origin}, for example, due to stellar activity. We find a difference of ($\Delta \ln{\mathcal{Z}} = 2.2$) in favour of the two-planet model compared to the model, including a quasi-periodic component for the \SI{15.5}{d} signal. This offers only weak evidence, confirming, nonetheless, that the signal is legitimately fitted by a Keplerian model.}

{Another} way to test the coherence of a signal for a given dataset is the use of the so-called stacked Bayesian {generalised} Lomb-Scargle periodogram \citep[s-BGLS;][]{Mortier.2015}. The diagram {in the left panel of \autoref{fig:bgls15d}} shows the probability of the \SI{15.5}{d} signal {normalised} to the minimum of the {considered} frequency range \citep{Mortier.2017} for an increasing number of observations. The period of the signal is uncertain when only a few observations are included, but starting with more than 80 observations, a signal of consistently rising probability is detected at the {period in question}. This indicates that the signal is stable in {phase and amplitude} over the whole observational period of a \SI{360}{\day} time baseline, as is likely for a planetary signal. {A colour-coded bar on the right side of the diagram specifies which of the instruments the considered data points originate from. As there are no {variations} of the signal caused by chunks of data from one {specific} instrument, we can {also} conclude that the signal is consistent between the different instruments.}

Even {though} there are no obvious signs of more than one transiting planet in the \emph{TESS} light curve (see \autoref{subsec:transit_search}), we searched for transits of the \SI{15.5}{d} signal based on its parameters {derived from the RV observations}. {Since the period of the planet is larger than half of the time span of the \emph{TESS} data, which comprise only one sector, a potential transit is likely to occur only once in the data.} {The right panel in \autoref{fig:bgls15d}} shows the {\emph{TESS} data phase-folded to the expected time of transit centre}. No obvious transit signals are visible. However, to quantify whether in fact there is no transit signal, we ran two more \texttt{juliet} fits on the \emph{TESS} data using Gaussian distributed priors based on the posterior of the planetary parameters in \autoref{tab:posterior_planets}. The model considering only the transiting planet is favoured by $\Delta\ln{\mathcal{Z}} \approx 3.6$ over the model that {treats the second periodicity as a transiting planet.} Thus, we conclude that no significant transiting signal is associated with the \SI{15.5}{d} periodicity.

\section{Discussion}
\label{sec:discussion}
\subsection{{GJ~3473}~b}
{\setlength{\extrarowheight}{4pt}%
\begin{table}
    \centering
    \caption{Derived planetary parameters of {GJ~3473}~b and c based on the posteriors of the joint fit.}
    \label{tab:derived_parameters}
    \begin{tabular}{lccl} 
        \hline
        \hline   
        Parameter & \multicolumn{2}{c}{Posterior\tablefootmark{(a)}} & Units \\
                       &                  {GJ~3473} b &    {GJ~3473} c  & \\
        \hline
        \noalign{\smallskip}
        \multicolumn{4}{c}{\textit{ Derived transit parameters}} \\
        \noalign{\smallskip}
        $p = R_{\rm p}/R_\star$            & $0.03184^{+0.00069}_{-0.00067}$  & \dots                         & \dots\\
        $b = (a/R_\star)\cos i_{\rm p}$    & $0.336^{+0.066}_{-0.074}$        & \dots                         & \dots\\
        $a/R_\star$                        & $9.39^{+0.19}_{-0.21}$           & \dots                         & \dots \\
        $i_p$                              & $87.95^{+0.47}_{-0.45}$          & \dots                         & deg \\
        $u_{1, TESS}$                      & $0.26^{+0.28}_{-0.18}$ \tablefootmark{(d)}   & \dots             & \dots \\
        $u_{2, TESS}$                      & $0.1{0}^{+0.28}_{-0.22}$            & \dots                         & \dots\\
        $t_T$                              & $0.95{0}^{+0.015}_{-0.014}$         & \dots                         & h \\
        \multicolumn{4}{c}{\textit{Derived physical parameters}\tablefootmark{(b)}} \\        
        $M_{\rm p}$                        & $1.86^{+0.30}_{-0.30}$           & $\geq 7.41^{+0.91}_{-0.86}$        & $M_\oplus$ \\
        $R_{\rm p}$                        & $1.264^{+0.05{0}}_{-0.049}$         & \dots                         & $R_\oplus$ \\
        $\rho_{\rm p}$                     & $5.03^{+1.07}_{-0.93}$           & \dots                         & \si{\gram\per\centi\meter\cubed} \\
        $g_{\rm p}$                        & $11.4^{+2.1}_{-2.0}$             & \dots                         & \si{\meter\per\second\squared} \\
        $a_{\rm p}$                        & $0.01589^{+0.00062}_{-0.00062}$  & $0.0876^{+0.0035}_{-0.0034}$  & \si{\astronomicalunit} \\
        $T_\textnormal{eq}$\tablefootmark{({c})}    & $773^{+16}_{-15}$         & $329.1^{+6.6}_{-6.4}$         & \si{\kelvin} \\
        $S$                                & $59.4^{+5.0}_{-4.5}$             & $1.95^{+0.17}_{-0.15}$        & $S_\oplus$ \\
        {ESM\tablefootmark{(d)}}    & {$6.8\pm0.3$}             & {\dots}                & {\dots} \\
        \hline
    \end{tabular}
    \tablefoot{
      \tablefoottext{a}{Error bars denote the $68\%$ posterior credibility intervals.}
      \tablefoottext{b}{We sample from a normal distribution for the stellar mass, stellar radius and stellar luminosity that is based on the results from \autoref{sec:stellar_prop}}.
      \tablefoottext{c}{Assuming a zero {Bond} albedo.}
      \tablefoottext{d}{{Emission spectroscopy metric \citep{Kempton.2018}}}}
\end{table}}

\begin{figure}
\centering
\includegraphics[]{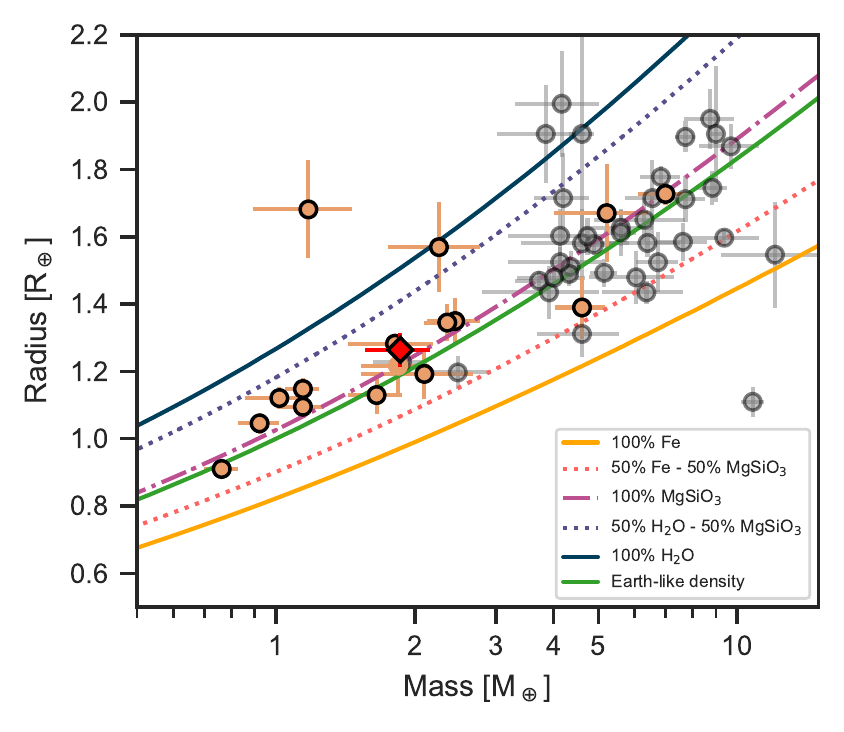}
\caption{Mass-radius diagram for small well {characterised} planets ($R<\SI{2}{R_\oplus}$, $\Delta M<\SI{30}{\percent}$) based on the TEPcat catalogue \citep[][visited on 14 April 2020]{Southworth.2011}. 
Planets orbiting stars with temperature $T_\mathrm{{eff}}<\SI{4000}{K}$ are displayed in orange colour, while the rest is displayed {as} grey circles. {GJ~3473}~b is marked {with} a red diamond. For comparison, theoretical mass-radius relations from \cite{Zeng.2016, Zeng.2019} are overlayed.}
\label{fig:mass_radius}
\end{figure}

\begin{figure}
\centering
\includegraphics[]{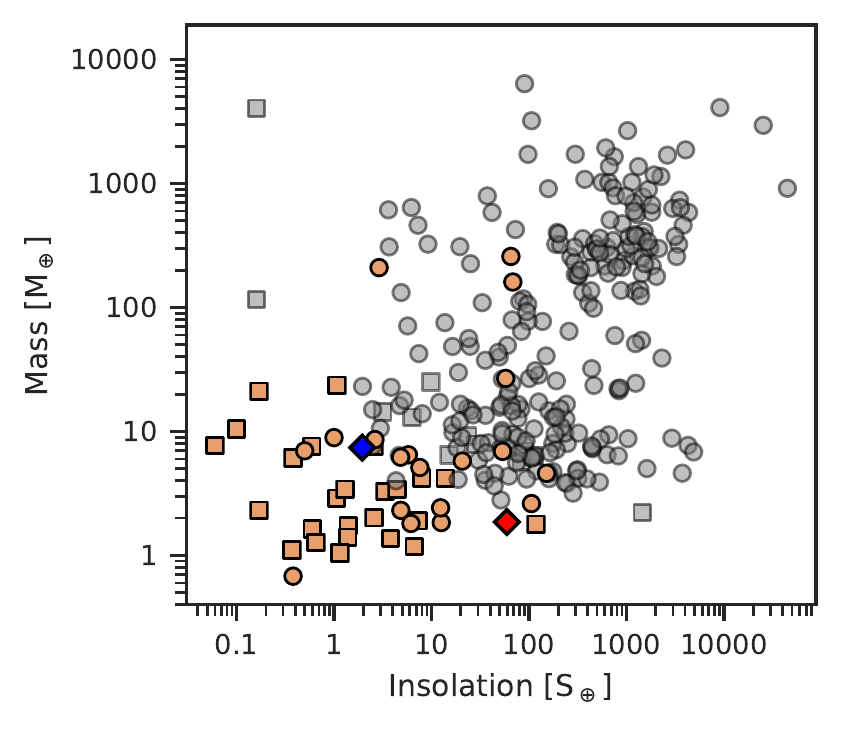}
\caption{{Mass-insolation diagram for small RV planets based {on the planetary systems composite data table of the \url{exoplanetarchive.ipac.caltech.edu/} (visited on 28 August 2020)}. 
Planets orbiting stars with temperature $T_\mathrm{{eff}}<\SI{4000}{K}$ are displayed in orange colour, while the rest is plotted {as} grey circles. Planet{s} with a {dynamical} mass measurement are shown as circles and planets with only a minimum mass {($M \sin{i}$)} measurement with boxes. {GJ~3473}~b and c are marked with red and blue diamonds.}}
\label{fig:mass_insolation}
\end{figure}

\begin{figure}
\centering
\includegraphics[]{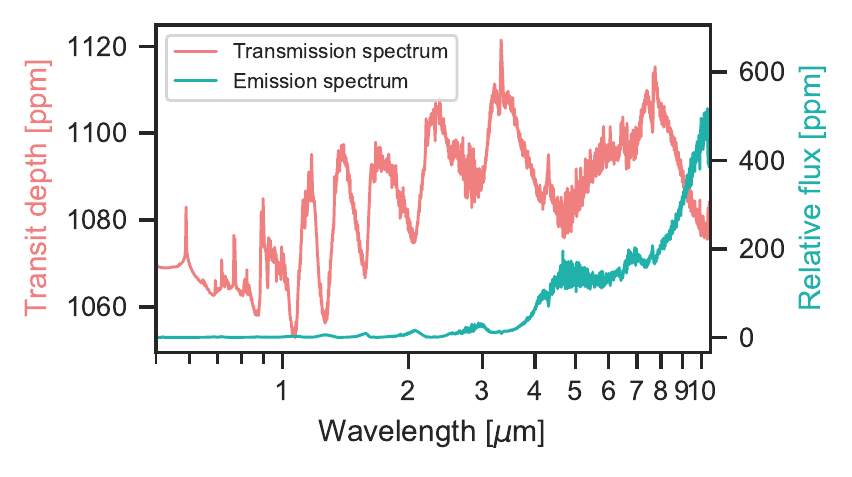}
\caption{{Representative} synthetic cloud free transmission and emission spectrum of {GJ~3473}~b.}
\label{fig:spectrum}
\end{figure}

Our derived mass and radius confirm the planetary nature of the transiting planet candidate detected by {\em TESS}. {GJ~3473}~b has {a} mass of $1.86^{+0.30}_{-0.30}$\,\si{M_\oplus} and a radius of $1.264^{+0.050}_{-0.049}$\,\si{R_\oplus}, which correspond to a density of $5.03^{+1.07}_{-0.93}$\,\si{\gram\per\centi\meter\cubed} and, thus, {fits} in the regime of Earth-sized planets with {a} density consistent with a {composition dominated by} MgSiO$_3$ (see \autoref{fig:mass_radius}). A summary of the derived physical parameters of the planet can be found in \autoref{tab:derived_parameters}.

{With an insolation flux of \SI{59.4\pm5}{S_\oplus}, {GJ~3473}~b is one of the hottest transiting Earth-mass planets with a {dynamical} mass measurement that has been detected so far (see \autoref{fig:mass_insolation}). Its equilibrium temperature corresponds to  \SI{773\pm16}{K}, assuming a zero {Bond} albedo. If the planet had an atmosphere,} thermochemical equilibrium calculations predict water and methane to be the dominant opacity sources in {the} near/mid infrared (NIR/MIR) of {the} transmission spectrum of {GJ~3473}~b, assuming a cloud-free solar-abundance scenario \citep[e.g.][]{Madhusudhan.2012, Molliere.2015, Molaverdikhani.2019}; see the red line in \autoref{fig:spectrum}. {In this scenario}, the main transmission spectral features in the optical are expected to be {alkali} (Na and K), although their {expected strength} depends on a number of parameters such as the planetary atmospheric metallicity. The emission spectrum is heavily muted by water and methane absorption, causing very low relative flux at wavelengths shorter than $\sim$3\,$\mu$m; see the blue line in \autoref{fig:spectrum}. The dominant spectral features of a cloudy atmosphere in the optical and NIR are expected to {be} similar to those of a cloud-free atmosphere, although with lower amplitudes and less pronounced methane features \citep{Molaverdikhani.2020}.

In addition, disequilibrium processes could change the composition and thermal structure of the planetary atmosphere. Depending on the exact temperature structure and methane abundance profile, vertical mixing could lead to methane quenching \citep[e.g.][]{Molaverdikhani.2019b}. Hydrocarbon haze (soot) production could act as a carbon-sink in the atmosphere, which might cause a reduced carbon-to-oxygen (C/O) ratio \citep[e.g.][]{Molaverdikhani.2019b, Gao.2020}. While haze opacities tend to obscure the optical to NIR wavelength {range}, {a} reduced C/O ratio could result in an enhancement of CO$_2$ production. This causes a prominent feature at around \SI{4.5}{\mu\meter} \citep[see e.g.][]{Kawashima.2019, Nowak.2020}. Atmospheres with higher metallicities {are} likely to lead to more prominent CO$_2$ features \citep[see e.g.][]{Heng.2016, Molaverdikhani.2019b, Nowak.2020, Schlecker.2020}. Hence, this spectral feature appears to be a key feature to retrieve planetary atmosphere metallicities, which, in turn, helps us to understand the formation history of the planet and the stellar system.

The amplitudes of the transmission spectral features of {GJ~3473}~b are estimated to be around \SIrange{10}{40}{ppm} {for the discussed model}. This {poses} a challenge for future observations of this planet through transmission spectroscopy. However, the relatively high temperature of this planet {causes} the emission spectral features at wavelengths longer than $\sim$3\,$\mu$m {to vary} from tens of ppm in NIR to hundreds ppm in MIR wavelengths up to \SI{4}{\mu\meter}. {We calculate the emission spectroscopy metric (ESM), based on \citet{Kempton.2018}, to be \num{6.8\pm0.3}, which is close to what \citet{Kempton.2018} classify as high-quality atmospheric characterisation targets ($\mathrm{ESM} > 7.5$).}

\subsection{{GJ~3473}~c}

Our RV modelling shows evidence for a second planet in the system. Its derived period is likely not linked to the stellar rotation period of \SI{168}{\day} {as} determined in \autoref{subsect:stellar_photometry}. Furthermore, the analysis of a comprehensive set of activity indicators exhibits no signs of stellar activity at the period in question. The analysis of the pEW of the H$\alpha$ line and the $\log{R'_\mathrm{HK}}$ index describes {GJ~3473} as a rather inactive star, which {would contradict} the relatively high RV amplitude of $\sim\SI{3.8}{\meter\per\second}$ if the signal was {attributed} to activity (cf. \autoref{subsec:activity}). Furthermore, the signal is coherent for at least one year of observations and invariant with respect to the different instruments (see \autoref{subsec:15dsignal}).

We therefore conclude that the \SI{15.5}{d} signal in the RVs is caused by a second planet in the system, {GJ~3473}~c. The planet {has} a lower mass limit of $7.41^{+0.91}_{-0.86}$\,\si{M_\oplus}. Further physical parameters derived for {this planet} are shown in \autoref{tab:derived_parameters}. No transit signals of {GJ~3473}~c are found within the {\emph{TESS}} data. An estimate of its bulk composition from theoretical models is not feasible because the derived mass places the planet just in the regime of the radius dichotomy between super-Earths and mini-Neptunes {\citep[e.g. {}][]{Owen.2013, Jin.2014, Fulton.2017, Zeng.2017, Cloutier.2020c}}. However, the non-detection of transits is not unexpected {when an orbit co-planar to {GJ~3473}~b ($i = \SI{87.95\pm0.47}{\degree}$) is assumed. The minimum inclination for at least grazing transits at a separation of \SI{0.0876\pm0.0035}{au} from the host star would be $i > \SI{89.47}{\degree}$ considering a planet at the empirical upper radius limit for mini-Neptunes.} At a distance of $8.66^{+0.13}_{-0.13}\times 10^{-2}$\,\si{\astronomicalunit} {from} the host star, {GJ~3473}~c receives $1.98^{+0.17}_{-0.15}$ times the {stellar flux compared to} Earth, which places it outside the inner boundary of the optimistic habitable zone, $1.49\,\si{S_\oplus} > \mathrm{S} > 0.22\,\si{S_\oplus}$, as defined by \cite{Kopparapu.2014}. {The planet therefore {is a} temperate super-Earth or mini-Neptune {such as} \object{GJ~887~c} \citep{Jeffers.2020}, \object{GJ~686~b} \citep{Lalitha.2019,Affer.2019}, \object{GJ~685} \citep{Pinamonti.2019} or \object{GJ~581~c} \citep{Udry.2007} (see \autoref{fig:mass_insolation}).}

\subsection{Comparison to synthetic planet populations}
We compare the planetary system of {GJ~3473} with a synthetic M dwarf planet population from a core accretion model of planet formation (Burn et al. in prep.) to assess the frequency of such a configuration. 
There, planets like {GJ~3473}~b are relatively abundant and often accompanied by multiple other planets in the system. 
More than \SI{10}{\percent} of their synthetic systems contain systems with a combination of planets similar to {GJ~3473}~b and c with respect to their masses and periods. 
The systems with an architecture closest to {GJ~3473} suggest a low bulk density for the outer planet, which can currently not be tested observationally.
Another theoretical prediction from the core accretion paradigm is a higher frequency of distant companions for volatile-poor inner planets {such as} {GJ~3473}~b \citep{Schlecker.2020}. While the current results do not demonstrate any clear evidence for planets beyond {GJ~3473}~c, further long-term monitoring is needed to {probe the outer system}.

\subsection{Search for transit timing variations}
\begin{figure*}
\centering
\includegraphics[]{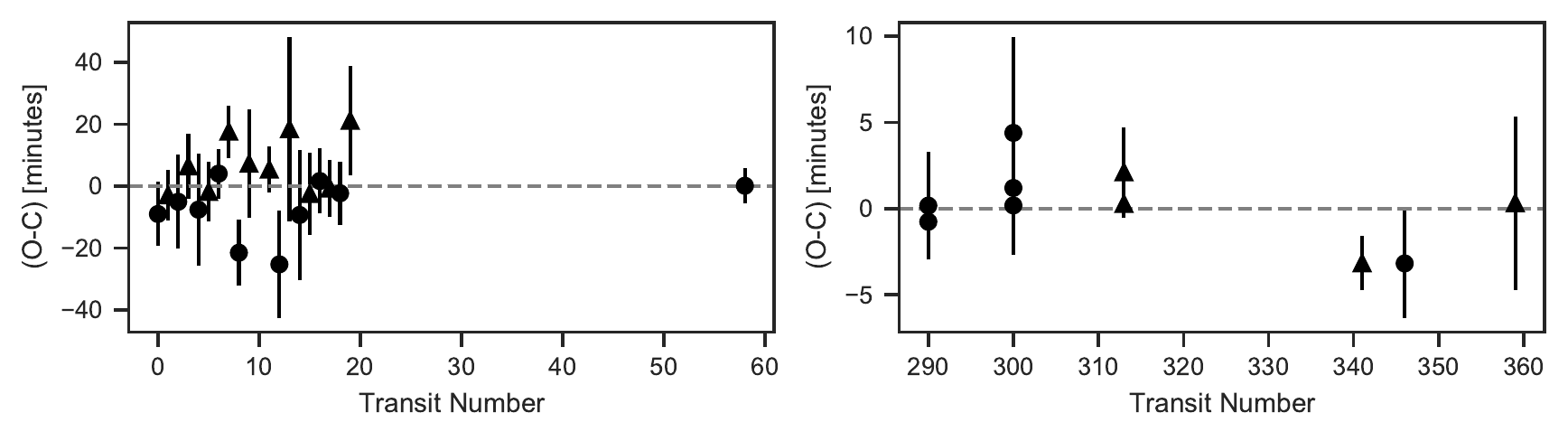}
\caption{TTVs measured for the transits of {GJ~3473}~b based on the results from the joint fit. Even transits are {depicted} as {circles} and odd transits as {triangles}. The observations corresponding to the transit numbers can be {found} in \autoref{tab:phot_follwup}.}
\label{fig:ttvs}
\end{figure*}

The period ratio of the two planets ($P_b\approx\SI{1.20}{d}$, $P_c\approx\SI{15.5}{d}$) does not suggest the presence of strong transit timing variations (TTV) for the transiting planet. However, we used \texttt{juliet} to {perform} a fit that only explores possible TTVs in the system. For this, we re-ran the joint fit but fixed all parameters to the results in \autoref{tab:posterior_planets} and \autoref{tab:posterior_instruments} and added a TTV parameter for each transit (Gaussian distributed prior with \num{0} mean and a standard deviation of \SI{0.03}{\day}, see the documentation of \texttt{juliet} for details). Although the results in \autoref{fig:ttvs} indicate TTVs up to $\sim \SI{20}{\minute}$, the error bars are rather large. The main reason for this is the small transit depth of {GJ~3473}~b compared to the scatter of the data points (see \autoref{fig:joint_fit_transit}). A GLS analysis of the TTVs reveals no significant periodicity that would indicate the presence of another massive planet in the system.

\section{Conclusions}
\label{sec:conclusions}

Here, we report the discovery of a planetary system around the M4.0\,V dwarf {GJ~3473} based on an extensive set of {RV} measurements from CARMENES, IRD, and HARPS, as well as space-based \emph{TESS} photometry and photometric transit follow-up observations from LCOGT, MuSCAT, and MuSCAT2, {and high-resolution images from Keck/NIRC2 and Gemini/NIRI}. We confirm the planetary nature of {GJ~3473}~b (TOI-488.01) and present its detailed {characterisation} from a simultaneous fit of the {RV} and transit data. The short-period planet has a mass of $M_b = \SI{1.86\pm0.30}{M_\oplus}$ and a radius of $R_b = \SI{1.264\pm0.050}{R_\oplus}$, which yields a density that is consistent with a rocky composition. The planet complements the sample of small planets with mass and radius measurements better than \SI{30}{\percent}  and contributes to the \emph{TESS} mission's primary goal to measure the masses of \num{50} planets with radii smaller than \SI{4}{R_\oplus}. Its proximity to the host star makes {GJ~3473}~b attractive for {thermal} emission spectroscopy. Synthetic cloud-free emission spectra predict amplitudes of the transmission spectral features up to 100s ppm in the MIR.

The {RV} data show evidence for an additional, non-transiting planet in the system. {GJ~3473}~c has a minimum mass of $M_c \sin{i} = \SI{7.41\pm0.91}{M_\oplus}$ and an orbital period of $P_c=\SI{15.509\pm0.033}{d}$, which places it just outside the inner boundary of the habitable zone.

{The planetary system of {GJ~3473} is another multi-planet system {discovered} around an M dwarf with planets in the range of Earth-like masses to super-Earths and mini-Neptunes.} {A} comparison with synthetic planet populations  shows that systems similar to {GJ~3473} may be relatively abundant and often host multiple planets. We therefore {encourage} further long-time monitoring of the system to find {additional} planets.

\begin{acknowledgements} 
CARMENES is an instrument at the Centro Astronómico Hispano-Alem\'an de Calar Alto (CAHA, Almería, Spain). 

CARMENES was funded by the German Max-Planck-Gesellschaft (MPG), the Spanish Consejo Superior de Investigaciones Científicas (CSIC), the European Union through FEDER/ERF FICTS-2011-02 funds, and the members of the CARMENES Consortium (Max-Planck-Institut für Astronomie, Instituto de Astrofísica de Andalucía, Landessternwarte Königstuhl, Institut de Ciències de l'Espai, Insitut für Astrophysik Göttingen, Universidad Complutense de Madrid, Thüringer Landessternwarte Tautenburg, Instituto de Astrofísica de Canarias, Hamburger Sternwarte, Centro de Astrobiología and Centro Astronómico Hispano-Alem\'an), with additional contributions by the Spanish Ministry of Economy, the German Science Foundation through the Major Research Instrumentation Programme and Deutsche Forschungsgemeinschaft (DFG) Research Unit FOR2544 ``Blue Planets around Red Stars'', the Klaus Tschira Stiftung, the states of Baden-Württemberg and Niedersachsen, and by the Junta de Andalucía. 
Part of this work was funded by the the Ministerio de Ciencia e Innovaci\'on via projects PID2019-109522GB-C51/2/3/4 and AYA2016-79425-C3-1/2/3-P, 
(MEXT/)JSPS KAKENHI via grants 15H02063, JP17H04574, 18H05442, JP18H01265, JP18H05439, 19J11805, JP19K14783, and 22000005, 
JST PRESTO via grant JPMJPR1775, {FCT/MCTES through national funds (PIDDAC, PTDC/FIS-AST/32113/2017) via grant UID/FIS/04434/2019,  FEDER - Fundo Europeu de Desenvolvimento Regional through COMPETE2020 - Programa Operacional Competitividade e Internacionalização (POCI-01-0145-FEDER-032113), the FONDECYT project 3180063, Funda\c{c}\~ao para a Ci\^encia e a Tecnologia through national funds and by FEDER through COMPETE2020 - Programa Operacional Competitividade e Internacionaliza\c{c}\~ao by these grants: UID/FIS/04434/2019; UIDB/04434/2020; UIDP/04434/2020; PTDC/FIS-AST/32113/2017 \& POCI-01-0145-FEDER-032113; PTDC/FIS-AST/28953/2017 \& POCI-01-0145-FEDER-028953.
} and the International Graduate Program for Excellence in Earth-Space Science.
Part of the data analysis was carried out on the Multi-wavelength Data Analysis System operated by the Astronomy Data Center, National Astronomical Observatory of Japan.
Funding for the {\emph{TESS}} mission is provided by NASA's Science Mission directorate.
We acknowledge the use of public {\emph{TESS}} Alert data from pipelines at the {\emph{TESS}} Science Office and at the {\emph{TESS}} Science Processing Operations Center.
This research has made use of the Exoplanet Follow-up Observation Program website, which is operated by the California Institute of Technology, under contract with the National Aeronautics and Space Administration under the Exoplanet Exploration Program.
Resources supporting this work were provided by the NASA High-End Computing Program through the NASA Advanced Supercomputing Division at Ames Research Center for the production of the SPOC data products.
This paper includes data collected by the {\emph{TESS}} mission, which are publicly available from the Mikulski Archive for Space Telescopes.
This work makes use of observations from the LCOGT network.
The analysis of this work has made use of a wide variety of public available software packages that are not referenced in the manuscript: \texttt{Exo-Striker} \citep{Trifonov.2019}, \texttt{astropy} \cite{AstropyCollaboration.2018},
\texttt{scipy} {\citep{Virtanen.2020}}, \texttt{numpy} \citep{Oliphant.2006}, \texttt{matplotlib} \citep{Hunter.2007}, \texttt{tqdm} \citep{daCostaLuis.2019}, \texttt{pandas} \citep{Thepandasdevelopmentteam.2020}, \texttt{seaborn} \citep{Waskom.2020}, \texttt{lightkurve} \citep{LightkurveCollaboration.2018} and \texttt{PyFITS} \citep{Barrett.2012}.
\end{acknowledgements}

\bibliographystyle{aa} 
\bibliography{references}

\begin{appendix} 

\onecolumn
\section{Known transiting planets with precise mass measurements around M dwarfs}
\begin{table*}[!h]
    \small
    \centering
    \caption{Small transiting planets with precise masses around M dwarfs.} 
    \label{tab:comparative_planets}
    \begin{tabular}{l
                    l
                    S[table-format=1.3]@{\,\( \pm \)\,}
                    S[table-format=1.3]
                    S[table-format=1.3]@{\,\( \pm \)\,}
                    S[table-format=1.3]
                    l}
        \hline
        \hline
        {Name}                          &        {Alternative name} & \multicolumn{2}{c}{Radius}       & \multicolumn{2}{c}{Mass}         & {Reference} \\
                                        &                           & \multicolumn{2}{c}{[R$_\oplus$]} & \multicolumn{2}{c}{[M$_\oplus$]} &             \\
        \hline
        {GJ~3473}~b$^{(a,b)}$                      & {G~50--16}~b                  & 1.264& 0.050                     & 1.86 & 0.30                      & This work \\
        \object{LP~729--54}~b$^{(a,b)}$  & LTT\,3780~b               & 1.35 & 0.06                      & 2.34 & 0.24                      & \cite{Nowak.2020}; \cite{Cloutier.2020} \\
        \object{TOI-1235}~b$^{(a,b)}$     & TYC\,4384--1735--1~b      & 1.69 & 0.08                      & 5.9 & 0.6                        & \cite{Bluhm.2020}; \cite{Cloutier.2020b} \\
        \object{GJ\,357}~b$^{(a,b)}$      & LHS\,2157~b               &1.217 &0.084                      & 1.84 & 0.31                      & \cite{Luque.2019}; \cite{Jenkins.2019} \\
        \object{GJ\,1252}~b$^{(a)}$     & L\,210--70~b              & 1.193 &0.074                     & 2.10 & 0.58                      & \cite{Shporer.2020} \\
        \object{L~98--59}~c$^{(a)}$      & TOI-175.01                & 1.35 &0.07                       & 2.42 & 0.35                      & \cite{Cloutier.2019}; \cite{Kostov.2019}  \\
        L~98--59~d$^{(a)}$               & TOI-175.01                & 1.57 &0.14                       & 2.31 & 0.46                      & \cite{Cloutier.2019}; \cite{Kostov.2019}    \\
        \object{L~168--9}~b$^{(a)}$      & CD--60\,8051~b              & 1.39 &0.09                       & 4.60 & 0.58                      & \cite{AstudilloDefru.2020} \\        
        \object{Kepler-138}~c           & KOI-314.2                 & 1.67 &0.15                       & 5.2 & 1.3                        & \cite{Almenara.2018}; \cite{Kipping.2014}; \cite{Mann.2017} \\
        Kepler-138~d                    & KOI-314.3                 & 1.68 &0.15                       & 1.17 & 0.30                      & \cite{Almenara.2018}; \cite{Kipping.2014}; \cite{Mann.2017} \\
        \object{GJ\,1132}~b             & LTT\,3758~b               & 1.130 &0.057                     & 1.66 & 0.23                      & \cite{Bonfils.2018}; \cite{BertaThompson.2015} \\
        \object{LHS\,1140}~b            & GJ\,3053~b                & 1.727 &0.033                     & 6.99 & 0.89                      & \cite{Ment.2019}; \cite{Dittmann.2017} \\
        LHS\,1140~c                     & GJ\,3053~c                & 1.282 &0.024                     & 1.81 & 0.39                      & \cite{Ment.2019} \\
        \object{TRAPPIST}-1~b           & 2MUCD 12171~b                     & 1.121 &0.033                     & 1.017 & 0.16                     & \cite{Grimm.2018}; \cite{Delrez.2018}; \cite{Gillon.2016}  \\
        TRAPPIST-1~c                    & 2MUCD 12171~c                     & 1.095 &0.031                     & 1.156 & 0.15                     & \cite{Grimm.2018}; \cite{Delrez.2018}; \cite{Gillon.2016}  \\
        TRAPPIST-1~d                    & 2MUCD 12171~d                     & 0.784 &0.023                     & 0.297 & 0.039                    & \cite{Grimm.2018}; \cite{Delrez.2018}; \cite{Gillon.2017} \\
        TRAPPIST-1~e                    & 2MUCD 12171~e                     & 0.910 &0.027                      & 0.772 & 0.079                    & \cite{Grimm.2018}; \cite{Delrez.2018}; \cite{Gillon.2017} \\
        TRAPPIST-1~f                    & 2MUCD 12171~f                     & 1.046 &0.030                     & 0.934 & 0.095                    & \cite{Grimm.2018}; \cite{Delrez.2018}; \cite{Gillon.2017}  \\
        TRAPPIST-1~g                    & 2MUCD 12171~g                     & 1.148 &0.033                     & 1.148 & 0.098                    & \cite{Grimm.2018}; \cite{Delrez.2018}; \cite{Gillon.2017}  \\
        TRAPPIST-1~h                    & 2MUCD 12171~h                     & 0.773 &0.027                     & 0.331 & 0.056                    & \cite{Grimm.2018}; \cite{Delrez.2018}; \cite{Luger.2017}  \\
        
        \hline        
    \end{tabular}
    \tablefoot{\tablefoottext{a}{Planets discovered by \emph{TESS}.}
    \tablefoottext{b}{Target stars in the CARMENES guaranteed time observations survey \citep{Quirrenbach.2014, Reiners.2018}.}
    
    The table is based on TEPCat \citep[][visited on 15 July 2020]{Southworth.2011} and shows the known transiting planets with radii smaller than \SI{2}{R_\oplus} and mass determinations to a precision better than \SI{30}{\percent} in orbits around stars with temperatures lower than \SI{4000}{\kelvin}. The first reference always denotes {the source of the properties}.}

\end{table*}

\newpage

\section{Priors for \texttt{juliet}}
\begin{table*}[!h]
    \centering
    \caption{Priors used for \texttt{juliet} in the joint fit of transits and RV.}
    \label{tab:priors_joint_fit}
    \begin{tabular}{llll} 
        \hline
        \hline
        \noalign{\smallskip}
        Parameter & Prior & Units & Description \\
        \noalign{\smallskip}
        \hline
        \noalign{\smallskip}

        \multicolumn{4}{c}{\textit{Stellar parameters}} \\
        \noalign{\smallskip}
        $\rho_\star$ & $\mathcal{N}(10520.0, 836.2)$ & \si{\kilo\gram\per\meter\cubed} & Stellar density \\
        \noalign{\smallskip}
        \multicolumn{4}{c}{\textit{Planetary parameters}} \\
        \noalign{\smallskip}
        $P_b$                    & $\mathcal{N}(1.1980004,0.000009)$    & d                        & Period of the transiting planet \\
        $t_{0,b}$                & $\mathcal{N}(2458492.2041,0.0015)$   & d                        & Time of transit centre of the transiting planet \\
        $r_{1,b}$                & $\mathcal{N}(0.55, 0.15)$            & \dots                    & Parametrisation for p and b \\
        $r_{2,b}$                & $\mathcal{N}(0.0318, 0.0021)$        & \dots                    & Parametrisation for p and b \\
        $K_{b}$                  & $\mathcal{N}(2.4,1.5)$               & $\mathrm{m\,s^{-1}}$     & Radial-velocity semi-amplitude of the transiting planet \\
        $\sqrt{e_{b}}\sin \omega_{b}$    & $\mathrm{fixed} (0)$         & \dots                    & Parametrisation for $e$ and $\omega$. \\
        $\sqrt{e_{b}}\cos \omega_{b}$    & $\mathrm{fixed} (0)$         & \dots                    & Parametrisation for $e$ and $\omega$.\\
        \noalign{\smallskip}
        $P_c$                    & $\mathcal{N}(15.51,0.16)$           & d                         & Period of the second RV signal \\
        $t_{0,c}$                & $\mathcal{N}(2458575.7,1.5)$        & d                         & Time of transit centre of the second RV signal \\
        $K_{c}$                  & $\mathcal{N}(3.7,1.5)$              & $\mathrm{m\,s^{-1}}$      & Radial-velocity semi-amplitude of the second RV signal \\
        $\sqrt{e_{c}}\sin \omega_{c}$    & $\mathrm{fixed}(0)$        & \dots        & Parametrisation for $e$ and $\omega$. \\
        $\sqrt{e_{c}}\cos \omega_{c}$    & $\mathrm{fixed}(0)$        & \dots            & Parametrisation for $e$ and $\omega$.\\     
        \multicolumn{4}{c}{\textit{Instrument parameters CARMENES, HARPS, IRD}} \\
        \noalign{\smallskip}
        $\mu$                     & $\mathcal{U}(-10,10)$              & $\mathrm{m\,s^{-1}}$      & Instrumental offset \\
        $\sigma$                  & $\mathcal{U}(0,10)$                & $\mathrm{m\,s^{-1}}$      & Jitter term \\
        \noalign{\smallskip}
        \multicolumn{4}{c}{\textit{Instrument parameters} \emph{TESS}} \\
        $q_{1}$                  & $\mathcal{U}(0,1)$                  & \dots                     & Quadratic limb-darkening parametrisation \\
        $q_{2}$                  & $\mathcal{U}(0,1)$                  & \dots                     & Quadratic limb-darkening parametrisation \\
        mdilution                & $\mathrm{fixed} (1)$                & \dots                     & Dilution factor \\
        mflux                    & $\mathcal{N}(0.0,.01)$               & ppm                       & Instrumental offset \\
        $\sigma$                 & $\mathcal{U}(1,500)$                & ppm                       & Jitter term \\
        \noalign{\smallskip}
        \multicolumn{4}{c}{\textit{Instrument parameters} MuSCAT2} \\
        $q_{1}$                  & $\mathcal{U}(0,1)$                  & \dots                     & Linear limb-darkening parametrisation \\
        mdilution                & $\mathrm{fixed} (1)$                & \dots                     & Dilution factor \\
        mflux                    & $\mathcal{N}(0.0,.01)$               & ppm                       & Instrumental offset \\
        $\sigma$                 & $\mathcal{U}(1,500)$                & ppm                       & Jitter term \\
        \noalign{\smallskip}
        \multicolumn{4}{c}{\textit{Instrument parameters} MuSCAT, LCOGT} \\
        $q_{1}$                  & $\mathcal{U}(0,1)$                  & \dots                     & Linear limb-darkening parametrisation \\
        mdilution                & $\mathrm{fixed} (1)$                & \dots                     & Dilution factor \\
        mflux                    & $\mathcal{N}(0.0,.01)$               & ppm                       & Instrumental offset \\
        $\sigma$                 & $\mathcal{U}(1,500)$                & ppm                       & Jitter term \\
        $\theta_0$               & $\mathcal{U}(-100, 100)$             & \dots                     & Linear airmass detrending coefficient \\        
        \hline
    \end{tabular}
    \tablefoot{The prior labels, $\mathcal{U}$ and $\mathcal{N,}$ represent uniform and normal distributions{, respectively}.}
\end{table*}

\begin{table*}
    \centering
    \caption{Priors used with \texttt{juliet} for the determination of the rotation period.}
    \label{tab:priors_rotation_period}
    \begin{tabular}{llll} 
        \hline
        \hline
        \noalign{\smallskip}
        Parameter & Prior & Units & Description \\
        \noalign{\smallskip}
        \hline
        \noalign{\smallskip}
        \multicolumn{4}{c}{\textit{Instrument parameters} Mearth, TJO} \\
        \noalign{\smallskip}
        mdilution                & $\mathrm{fixed} (1)$                & \dots                     & Dilution factor \\
        mflux                    & $\mathcal{N}(0.0, 1e5)$               & ppm                       & Instrumental offset \\
        $\sigma$                 & $\mathcal{J}(1e-5,1e5)$                & ppm                       & Jitter term \\
        \multicolumn{4}{c}{\textit{GP parameters (individual)} Mearth, TJO} \\
        \noalign{\smallskip}        
        GP-$\sigma$              & $\mathcal{J}(1e-8,1e8)$              & ppm                       & GP amplitude \\
        GP-$\Gamma$              & $\mathcal{J}(1e-2,1e2)$              & \dots                     & GP amplitude of the sine-squared component \\
        \multicolumn{4}{c}{\textit{GP parameters (shared)} Mearth, TJO} \\
        \noalign{\smallskip} 
        GP-$\alpha$              & $\mathcal{J}(1e-10,1)$              & d$^{-2}$                    & GP inverse length scale of the exponential component \\
        GP-$P_{\mathrm{rot}}$    & $\mathcal{U}(2,200)$                & d                           & GP rotation period of the quasi-periodic component \\
        \hline
    \end{tabular}
    \tablefoot{The prior labels, $\mathcal{U}$ and $\mathcal{N,}$ represent uniform, and normal distributions. $\mathcal{J}$ is the log-uniform Jeffrey's distribution \citep{Jeffreys.1946}.}
\end{table*}

\clearpage

\section{Continuation of the posteriors}
{\setlength{\extrarowheight}{4pt}
    \begin{table*}[!h]
    \centering
    \caption{Posteriors of the joint fit for the different instrumental parameters.}
    \label{tab:posterior_instruments}
    \begin{tabular}[t]{l c l}
        \hline
        \hline
        \noalign{\smallskip}
        Parameter & Posterior\tablefootmark{(a)} & Units \\
        \noalign{\smallskip}
        \hline
        \noalign{\smallskip}
        \multicolumn{3}{c}{\textit{TESS}} \\
        $q_1$ &  $0.17^{+0.25}_{-0.12}$ &  \dots \\
        $q_2$ &  $0.35^{+0.33}_{-0.24}$ &  \dots \\
        $\sigma$ &  $34^{+31}_{-22}$ &  ppm \\
        \noalign{\smallskip}
        \multicolumn{3}{c}{\textit{LCO McD$_{z_s}$ 19 Mar. 2019}} \\
        $q_1$ &  $0.49^{+0.29}_{-0.30}$ &  \dots \\
        $\sigma$ &  $437^{+42}_{-78}$ &  ppm \\
        $\theta_0$ &  $-0.00302^{+0.00058}_{-0.00058}$ &  \dots \\
        \noalign{\smallskip}
        \multicolumn{3}{c}{\textit{MuSCAT2$_i$ 21 Dec. 2019}} \\
        $q_1$ &  $0.77^{+0.16}_{-0.26}$ &  \dots \\
        $\sigma$ &  $113^{+104}_{-72}$ &  ppm \\
        \noalign{\smallskip}
        \multicolumn{3}{c}{\textit{MuSCAT2$_{zs}$ 21 Dec. 2019}} \\
        $q_1$ &  $0.57^{+0.26}_{-0.31}$ &  \dots \\
        $\sigma$ &  $155^{+136}_{-100}$ &  ppm \\
        \noalign{\smallskip}
        \multicolumn{3}{c}{\textit{MuSCAT2$_i$ 2 Jan. 2020}} \\
        $q_1$ &  $0.57^{+0.27}_{-0.31}$ &  \dots \\
        $\sigma$ &  $114^{+99}_{-72}$ &  ppm \\
        \noalign{\smallskip}
        \multicolumn{3}{c}{\textit{MuSCAT2$_r$ 2 Jan. 2020}} \\
        $q_1$ &  $0.54^{+0.28}_{-0.31}$ &  \dots \\
        $\sigma$ &  $121^{+109}_{-77}$ &  ppm \\
        \noalign{\smallskip}
        \multicolumn{3}{c}{\textit{MuSCAT2$_{zs}$ 2 Jan. 2020}} \\
        $q_1$ &  $0.78^{+0.15}_{-0.23}$ &  \dots \\
        $\sigma$ &  $109^{+90}_{-69}$ &  ppm \\
    \hline
    \end{tabular}\quad\quad\quad\quad
    \begin{tabular}[t]{l c l}    
        \hline
        \hline
        \noalign{\smallskip}
        Parameter & Posterior\tablefootmark{(a)} & Units \\
        \noalign{\smallskip}
        \hline         
         \noalign{\smallskip}
        \multicolumn{3}{c}{\textit{MuSCAT$_r$ 18 Jan. 2020}} \\
        $q_1$ &  $0.867^{+0.09}_{-0.172}$ &  \dots \\
        $\sigma$ &  $385^{+76}_{-143}$ &  ppm \\
        $\theta_0$ &  $0.0054^{+0.0011}_{-0.0011}$ &  \dots \\
        \noalign{\smallskip}
        \multicolumn{3}{c}{\textit{MuSCAT$_zs$ 18 Jan. 2020}} \\
        $q_1$ &  $0.147^{+0.150}_{-0.098}$ &  \dots \\
        $\sigma$ &  $440.0^{+38.0}_{-55.0}$ &  ppm \\
        $\theta_0$ &  $0.00346^{+0.00067}_{-0.00069}$ &  \dots \\
        \noalign{\smallskip}
        \multicolumn{3}{c}{\textit{LCO CTIO$_{ip}$ 21 Feb. 2020}} \\
        $q_1$ &  $0.873^{+0.085}_{-0.158}$ &  \dots \\
        $\sigma$ &  $492.9^{+4.9}_{-9.2}$ &  ppm \\
        $\theta_0$ &  $0.0015^{+0.00017}_{-0.00017}$ &  \dots \\
        \noalign{\smallskip}
        \multicolumn{3}{c}{\textit{LCO CTIO$_{ip}$ 27 Feb. 2020}} \\
        $q_1$ &  $0.52^{+0.28}_{-0.30}$ &  \dots \\
        $\sigma$ &  $482^{+12}_{-22}$ &  ppm \\
        $\theta_0$ &  $0.00243^{+0.00017}_{-0.00018}$ &  \dots \\
        \noalign{\smallskip}
        \multicolumn{3}{c}{\textit{LCO SAAO$_{zs}$ 13 Mar. 2020}} \\
        $q_1$ &  $0.60^{+0.26}_{-0.33}$ &  \dots \\
        $\sigma$ &  $413^{+56}_{-98}$ &  ppm \\
        $\theta_0$ &  $-0.00086^{+0.0007}_{-0.00071}$ &  \dots \\
    \hline
    \end{tabular}
    \tablefoot{\tablefoottext{a}{Error bars denote the $68\%$ posterior credibility intervals.}}
\end{table*}}

\end{appendix}

\end{document}